\documentclass[aip, pop, reprint, twocolumn]{revtex4-2}
\usepackage{amsmath,amssymb}
\usepackage{graphicx}
\usepackage{dcolumn}
\usepackage{bm}

\usepackage[utf8]{inputenc}
\usepackage[T1]{fontenc}
\usepackage{newtxtext,newtxmath}

\usepackage{etoolbox}
\usepackage{xcolor}
\usepackage{graphicx}
\usepackage{subcaption}
\usepackage{caption}
\usepackage{xcolor,soul} 
\sethlcolor{yellow} 
\usepackage{hyperref}

\usepackage{CJKutf8}

\makeatletter
\def\@email#1#2{%
 \endgroup
 \patchcmd{\titleblock@produce}
  {\frontmatter@RRAPformat}
  {\frontmatter@RRAPformat{\produce@RRAP{*#1\href{mailto:#2}{#2}}}\frontmatter@RRAPformat}
  {}{}
}%

\makeatother
\begin{document}
\begin{CJK}{UTF8}{min}


\title{\textit{A Priori} Assessment of Rotational Invariance in Multiscale Convolutional Neural Network-Based Subgrid-Scale Model for Wall-Bounded Turbulent Flows}

\author{Bahrul Jalaali$^*$}
    \affiliation{ 
Department of Mechanical Engineering, The University of Osaka, 2-1 Yamadaoka, Suita, Osaka 565-0871, Japan
}%
\author{Kie Okabayashi(岡林 希依)}%
    \affiliation{ 
Department of Mechanical Engineering, The University of Osaka, 2-1 Yamadaoka, Suita, Osaka 565-0871, Japan
}%
\email{bahrul.jalaali@mail.ugm.ac.id} 

\date{\today}

\begin{abstract}

This study proposes a rotationally invariant data-driven subgrid-scale (SGS) model for large-eddy simulation (LES) of wall-bounded turbulent flows. Building upon the multiscale convolutional neural network subgrid-scale model, which outputs SGS stress tensors ($\tau_{ij}$) as the baseline, the deep neural network (DNN) architecture is modified to satisfy the principle of material objectivity by removing the bias terms and batch normalization layers while incorporating a spatial transformer network (STN) algorithm. The model was trained on a turbulent channel flow at $\mathrm{Re}_\tau = 180$ and evaluated using both non-rotated and rotated inputs. The results show that the model performs well in predicting $\tau_{ij}$ and key turbulence statistics, including dissipation, backscatter, and SGS transport. These quantities reflect the ability of the model to reproduce the energy transfer between the resolved scale and SGS. Moreover, it effectively generalizes to unseen rotated inputs, accurately predicting $\tau_{ij}$ despite the input configurations not being encountered during the training. These findings highlight that modifying the DNN architecture and integrating the STN-based algorithm improves the ability to recognize and correctly respond to rotated inputs. The proposed data-driven SGS model addresses the key limitations of common data-driven SGS approaches, particularly their sensitivity to rotated input conditions. It also marks an important advancement in data-driven SGS modeling for LES, particularly in flow configurations where rotational effects are non-negligible.
\end{abstract}

\maketitle

\section{Introduction}

Large-eddy simulation (LES) is a numerical approach for accurately capturing the dynamics of turbulent flows in both engineering and science applications. In contrast to direct numerical simulation (DNS), which resolves all scales of motion, LES resolves large-scale or grid-scale (GS) dynamics and represents the small-scale or subgrid-scale (SGS) structures using the SGS model. Conventional SGS models, especially eddy-viscosity models (EVM), have the advantages of simplicity and robustness. However, the inherent linearity of an EVM may limit its accuracy, especially in capturing anisotropic turbulence\cite{spalartphilosophies2015}. Efforts made to address this limitation include the scale-similarity (SS) model \cite{bardinaimproved1980} and nonlinear model or Clark model \cite{clarkevaluation1979}. In the SS model, the SGS structures are assumed to be similar across neighboring scales, meaning that the flow structures observed at different filtering levels are similar to one another. This model employs a test filter to approximate the SGS stresses ($\tau_{ij}$) through the modified Leonard tensor\cite{bardinaimproved1980}. The nonlinear model is based on SGS stresses explicitly derived from a Taylor-series expansion of the filtering operation. Both models reconstruct the SGS stresses directly without invoking the eddy-viscosity assumption. Although these models are well-correlated with DNS data, they tend to lack sufficient dissipation\cite{kajishima2017,sagaut2006}. Advancements such as the mixed model\cite{Bardina1983}, which combines the SS model with Smagorinsky model\cite{smag1963}, and the dynamic Smagorinsky model (DSM)\cite{germano}, have been proposed. The mixed model leverages the local structural accuracy of the SS model and introduces a dissipation term from the Smagorinsky model. This model remains constrained by the assumptions underlying each of the SS and Smagorinsky components\cite{kajishima2017,sagaut2006}. Germano et al.\cite{germano} proposed the SGS model for adapting the Smagorinsky model by dynamically adjusting the Smagorinsky constant at each point in space and time to improve the adaptability of the model to the local flow structure. The computation of this constant aims to avoid \textit{a priori} estimation of the Smagorinsky model by using the Germano identity with both the grid filter and an additional test filter\cite{germano}. Nevertheless, this model suffers from instability owing to negative coefficients, and numerical averaging is required to achieve stability\cite{sagaut2006,kajishima2017}. However, this averaging offsets the main advantage of the DSM to determine the local and dynamic Smagorinsky constants directly from the resolved flow field\cite{kajishima2017}.

Owing to the challenges faced by conventional SGS models, data-driven strategies have recently gained significant attention as an alternative approach for reconstructing SGS models. Deep neural networks (DNN), trained on spatially filtered DNS (fDNS) datasets, have demonstrated the ability to learn mappings from resolved flow variables to the unclosed $\tau_{ij}$. In contrast to conventional SGS models, data-driven approaches aim to inductively extract small-scale SGS features without relying on artificial assumptions. To the best of the authors’ knowledge, the earliest application of a data-driven SGS model was proposed by Sarghini et al.~\cite{Sarghini2003}, who estimated the eddy-viscosity coefficient using a multilayer perceptron (MLP) instead of directly calculating $\tau_{ij}$ in the momentum equation. Their results were comparable to those of the conventional Bardina model in terms of the turbulence statistics\cite{Sarghini2003}. Subsequently, in contrast to the approach of Sarghini et al. \cite{Sarghini2003}, who employed a shallow DNN, Pal \cite{pal2019deeplearningparameterizationsubgrid} developed an MLP-based SGS model with a deeper DNN architecture to reconstruct the eddy-viscosity coefficient. This deeper architecture allowed the model to represent more complex mappings between the resolved flow variables and $\tau_{ij}$, thereby improving its ability to capture the turbulence features. As that of result, their findings demonstrated improved performance over the DSM in channel-flow simulations, with the turbulence statistics demonstrating a closer agreement with the DNS data\cite{pal2019deeplearningparameterizationsubgrid}. Yet, the effectiveness of these approaches is often attributed to their dependence on the alignment with the rate-of-strain tensor $\bar{D}_{ij}$, as the accuracy hinges on the eddy-viscosity formulation, which is a form of artificial assumption\cite{Pope2000}. Here, the overbar $\overline{(\cdot)}$  denotes the filtering operation.

Rather than inferring the eddy-viscosity coefficients, researchers have adopted a direct-closure strategy by predicting $\tau_{ij}$ directly from the resolved turbulent fields (e.g., Refs.~\cite{beck2019, kurz2022, gamahara2017, zhou2019, liu2022, xie2020, park2021, guan2022, jalaali2025}). These approaches envision functions that are not constrained by artificial assumptions, such as an EVM, thus fully leveraging the benefits of data-driven models. Beck et al.~\cite{beck2019} suggested that the data-driven SGS model is feasible as long as the direct-closure terms are obtained from a previously conducted DNS. Accordingly, Kurz $\&$ Beck~\cite{kurz2022} employed an MLP-based SGS model and showed a good agreement with the fDNS for isotropic turbulence. A similar approach has been adopted by several researchers, such as Gamahara and Hattori \cite{gamahara2017}, Zhou et al. \cite{zhou2019}, and Liu et al. \cite{liu2022}, who predicted $\tau_{ij}$ using resolved flow variables as the input to their MLP-based SGS model. Their results demonstrated that this direct-closure strategy could reproduce the SGS stress distribution in \textit{a priori} test and turbulence statistics in \textit{a posteriori} test with a close agreement with the result of the fDNS. 


In subsequent studies, instead of applying an MLP algorithm in which both input and output are defined at single-point locations, researchers such as Xie et al. \cite{xie2020} and Park $\&$ Choi \cite{park2021} have explored multipoint inputs to improve the predictive accuracy. This approach enables the MLP algorithm to capture turbulent interactions by accounting for non-local input features. Xie et al. \cite{xie2020} used multipoint data of the velocity gradient $\frac{\partial \bar{u}_i}{\partial x_j}$ as the input and achieved accurate \textit{a priori} predictions, along with improvements in \textit{a posteriori} tests when compared with the results of the conventional DSM for isotropic turbulence. Park $\&$ Choi \cite{park2021} evaluated both single- and multipoint inputs in a channel flow case. They found that multipoint inputs improved the $\tau_{ij}$ prediction accuracy and achieved a good agreement with the fDNS result in \textit{a posteriori} test. However, it is worth noting that MLP-based algorithms with multiple inputs often impose a substantial computational burden, making them less effective for large-scale datasets\cite{liu2022}. 

A more efficient approach to extract multipoint spatial information is through the use of convolutional neural networks (CNN), which capture the spatial correlations via convolutional kernels such that both the memory and computational cost can be reduced, making them more suitable for flow-field data\cite{morimoto2021,liu2022}. Given these advantages, CNN-based SGS models have been employed in previous studies, such as those of Beck et al. \cite{beck2019} and Guan et al. \cite{guan2022}, for homogeneous isotropic turbulence. Their results demonstrated accurate predictions of $\tau_{ij}$ and showed that the CNN-based models outperformed both static and dynamic Smagorinsky models\cite{beck2019,guan2022} and the MLP-based SGS models in \textit{a posteriori} test. Liu et al. \cite{liu2022} employed a CNN-based data-driven SGS model for a turbulent channel flow. They also compared the CNN-based and MLP-based SGS models, showing that the CNN-based model provided more accurate predictions of $\tau_{ij}$. In \textit{a posteriori} tests, the CNN-based model outperformed the DSM in terms of turbulence statistics. This improvement underscores the advantage of using CNN to capture spatial and turbulent interactions across the neighboring grid points. Furthermore, it is crucial to represent the multiscale characteristics of turbulent flow dynamics\cite{xie2020, jalaali2025}. Xie et al.\cite{xie2020} suggested that data-driven SGS models should capture the multiscale characteristic and energy cascade of turbulence through appropriately designed multiple-point input features. To this end, Jalaali and Okabayashi \cite{jalaali2025} introduced a multiscale CNN-based SGS model designed to capture the vortex interactions across scales, which are essential for energy and momentum transfer. Their model achieved more accurate predictions of $\tau_{ij}$ and improved the turbulent statistics in \textit{a posteriori} tests when compared with those of both the conventional Smagorinsky model and other CNN-based data-driven SGS models. Moreover, it reproduced the inverse energy cascade and maintained numerical stability in \textit{a posteriori} tests\cite{jalaali2025}. Such improvements are likely attributable to the ability of the model to capture interscale vortex interactions, which are not explicitly represented in conventional CNN-based approaches.
 
Despite these promising developments, for any data-driven SGS model to be physically valid, it must adhere to fundamental invariance principles inherent to turbulence modeling. In modeling terms, the turbulence model must be consistent and frame-indifferent, or satisfy the principle of material objectivity. Fulfilling this requirement has been treated as a hard constraint not only for conventional SGS models\cite{kajishima2017} but also the data-driven SGS models \cite{Spalart2023,duraisamy2021}. As discussed by Oberlack\cite{Oberlack1997}, classical SGS models often fail to meet this requirement. In particular, models that explicitly include the rate-of-rotation tensor frequently violate the requirement of frame indifference. To address this, considerable effort has been made to incorporate invariance into the SGS formulations. For example, Oberlack\cite{Oberlack1997} noted that the DSM proposed by Germano et al.\cite{germano} satisfied this principle by introducing invariant variables and test-filter quantities that preserved the scaling invariance. Although the model structures considerably differ from those of the conventional SGS models, data-driven SGS models must also adhere to material objectivity, ensuring physically consistent behavior under coordinate rotations. Accordingly, data-driven SGS models are expected to respect the following transformation property:
\begin{equation}
\mathcal{F}(R \bar{X} R^{T}) = R \mathcal{F}(\bar{X}) R^{T},
\label{Eq1}
\end{equation}
where $\mathcal{F}$ denotes the learned mapping from the filtered input features $\bar{X}$. The matrix $R$ is an arbitrary rigid-body rotation operator representing a transformation of the coordinate frame.

Likewise, the efforts to satisfy the material objectivity have been undertaken within the data-driven turbulence modeling framework. In this regard, the generalized expansion of the Reynolds stress tensor by Pope \cite{pope1975} has been employed to develop invariant closures for Reynolds stresses. This approach is based on the assumption that the stresses depend on the mean velocity gradient \cite{Pope2000}. Within the data-driven framework, this model rewrites the expansion of the Reynolds stress tensor by Pope as

\begin{equation}
{\tau_{ij}} = \sum_{n=1}^{10} c^{(n)}(\lambda_{1}, \ldots, \lambda_{5}),{T}^{(n)},
\label{tbnn}
\end{equation}
where $T^{(n)}$ are known functions of the symmetric and antisymmetric parts of the velocity gradient tensor, and $\lambda_{1}, \ldots, \lambda_{5}$ are scalar invariants. The detailed derivation of these 10 tensors and 5 invariants is given in Pope\cite{Pope2000,pope1975}. The goal is to determine the corresponding coefficients ($c^{(n)}(\lambda_{1}, \ldots, \lambda_{5})$), following which the Reynolds stress tensor can be calculated using Eq.~\eqref{tbnn}. The data-driven approach was initially introduced by Ling et al. \cite{Lingmachine2016, LingReynolds2016} as the tensor basis neural network (TBNN), and later extended by other researchers, such as Brener et al. \cite{brener2024} in the context of data-driven RANS modeling and Bose $\&$ Roy\cite{bose2024} for LES closures. These studies relied on scalar invariants to reconstruct the anisotropic stresses and showed a relatively good agreement with the fDNS data.

However, there are limitations to the TBNN approach. As noted by Cinnella \cite{cinnella2024}, the TBNN based on the representation by Pope relies on the hypothesis of the local equilibrium of turbulence, an assumption that no longer holds for flows with strong non-equilibrium. To some extent, the polynomial expansion used in the representation by Pope can be extended to better capture the non-equilibrium turbulence\cite{pope1975,wu1028}. Prakash et al. \cite{prakash2022} noted that the size of the minimal tensor set and the corresponding invariant basis increases with the number of prescribed input features. To address this issue, Prakash et al. \cite{prakash2022} proposed a data-driven SGS model based on the eigenstructure of the SGS tensor, using invariant variables as inputs. Their results showed a good agreement with the DNS data for isotropic turbulence. Brener et al. \cite{brener2024} and Wu et al.\cite{wu1028} showed that invariance can also be achieved through data augmentation. By exposing the DNN architecture to multiple orientations of the dataset, the data-driven SGS model was expected to learn representations that are insensitive to frame rotations\cite{wu1028}. However, data augmentation was computationally expensive because it required duplicating the DNS datasets across multiple transformed coordinate systems\cite{brener2024}. Beyond input design and data augmentation, a critical concern remains that invariance should be preserved not only at the level of the input features but also within the structure and operations of the DNN algorithm itself\cite{Jaderberg, prakash2022, bose2024}. For example, the study by Jalaali $\&$ Okabayashi\cite{jalaali2025} used rotationally invariant inputs, but the architecture of the DNN model itself did not maintain rotational invariance as, a CNN algorithm cannot preserve rotational invariance without the necessary treatment\cite{goodfellow2016}. Because DNN is inherently composed of coordinate-sensitive operations, it is desirable to design architectures or training strategies that are intrinsically invariant. 

Motivated by these challenges, we developed a data-driven SGS model that enforces material objectivity and preserves rotational invariance. Here, we adopted the multiscale CNN-based SGS model introduced by Jalaali $\&$ Okabayashi \cite{jalaali2025} (MSC model), which captures the flow features across multiple spatial scales. The contributions of this study can be outlined as follows. First,  we modified the MSC model by removing the bias terms and batch normalization (BN) layers, following Bin et al. \cite{Bin2022} and Cho et al. \cite{Cho2024}, allowing the model to operate directly on raw input data and enhancing the invariance properties. Second, we integrated a spatial transformer network (STN) \cite{Jaderberg} into the architecture, enabling the model to learn the features that remain consistent under coordinate rotations. These design choices were evaluated through \textit{a priori} test with rotated inputs to examine the rotational invariance. The remainder of this paper is organized as follows. In Section II, we introduce the numerical framework and present the proposed data-driven SGS model, with particular emphasis on the overall modeling strategy. Section III describes the results and discussions. Finally, Section IV concludes the study by summarizing the key findings and discussing possible directions for future developments.

\section{Methodology}
    \subsection{\label{sec:level2}Problem setting}  
    This study investigated a wall-bounded turbulent channel flow between two parallel flat plates, driven by a constant pressure gradient. The governing equations for the LES are obtained by solving the spatially filtered forms of the continuity and Navier$\--$Stokes equations, as follows:

    \begin{equation}
        \frac{\partial \bar{u}_i}{\partial x_i} = 0.
        \label{mass}
    \end{equation}
    
    \begin{equation}
        \frac{\partial \bar{u}_i}{\partial t} + \frac{\partial (\bar{u}_i \bar{u}_j)}{\partial x_j} = - \frac{\partial \bar{p}}{\partial x_i} + \frac{1}{\text{Re}_{\tau}} \frac{\partial}{\partial x_j} \left(-\tau_{ij} + 2\bar{D}_{ij}\right).
        \label{momentum}
    \end{equation}
    
    Here, $\text{Re}_{\tau}$ is the frictional Reynolds number, and $\bar{D}_{ij}$ denotes the GS rate-of-strain tensor, defined as:
    
    \begin{equation}
        \bar{D}_{ij} = \frac{1}{2} \left( \frac{\partial \bar{u}_i}{\partial x_j} + \frac{\partial \bar{u}_j}{\partial x_i} \right).
        \label{dij}
    \end{equation}

\begin{figure}
    \includegraphics[width=1\linewidth]{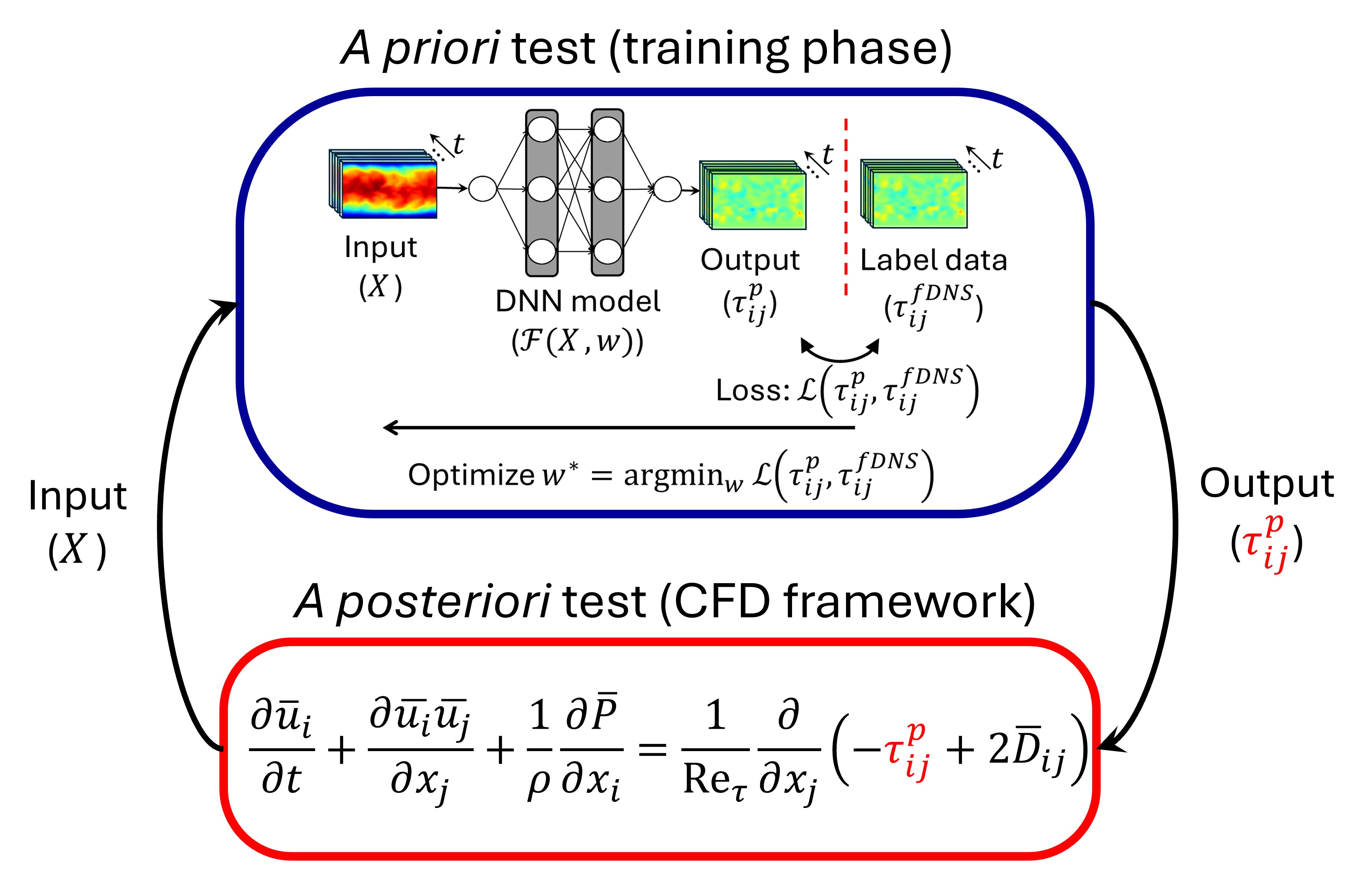}
    \caption{\label{Fig1} Schematic of data-driven SGS model framework.}
    \end{figure}

    The residual SGS stress, $\tau_{ij} = \overline{u_i u_j} - \bar{u}_i \bar{u}_j$, is unclosed and represents the effect of unresolved scales. Therefore, a closure model is required; while by far the most widely used closure adopts the eddy-viscosity hypothesis that can be modeled using conventional approaches such as the Smagorinsky model\cite{smag1963}. In the present study, we replaced such conventional closures with a data-driven SGS model for $\tau_{ij}$. Fig.~\ref{Fig1} outlines the modeling workflow. This study was limited to \textit{a priori} assessment of the capability of the proposed data-driven SGS model to achieve rotational invariance. In \textit{a priori} tests, the data-driven SGS model is trained in a supervised manner on fDNS flow fields. By using the input data ${X}$ and label data $\tau_{ij}$, the data-driven SGS model learns a nonlinear mapping such that $\tau_{ij}^{p} = \mathcal{F}({X}; w)$, where $\tau_{ij}^{p}$ is the predicted $\tau_{ij}$. The parameter $w$ denotes the weights of the DNN and is obtained by optimization, given as $w = \arg \min_w \left( \mathcal{L}(\tau_{ij}, \tau_{ij}^{p}) \right)$. $\mathcal{L}$ denotes the loss function.
    
    \subsection{\label{sec:level2}Dataset preparation}
\begin{figure*} 
    \centering
        \begin{subfigure}{0.45\textwidth} 
            \centering
            \includegraphics[width=\linewidth]{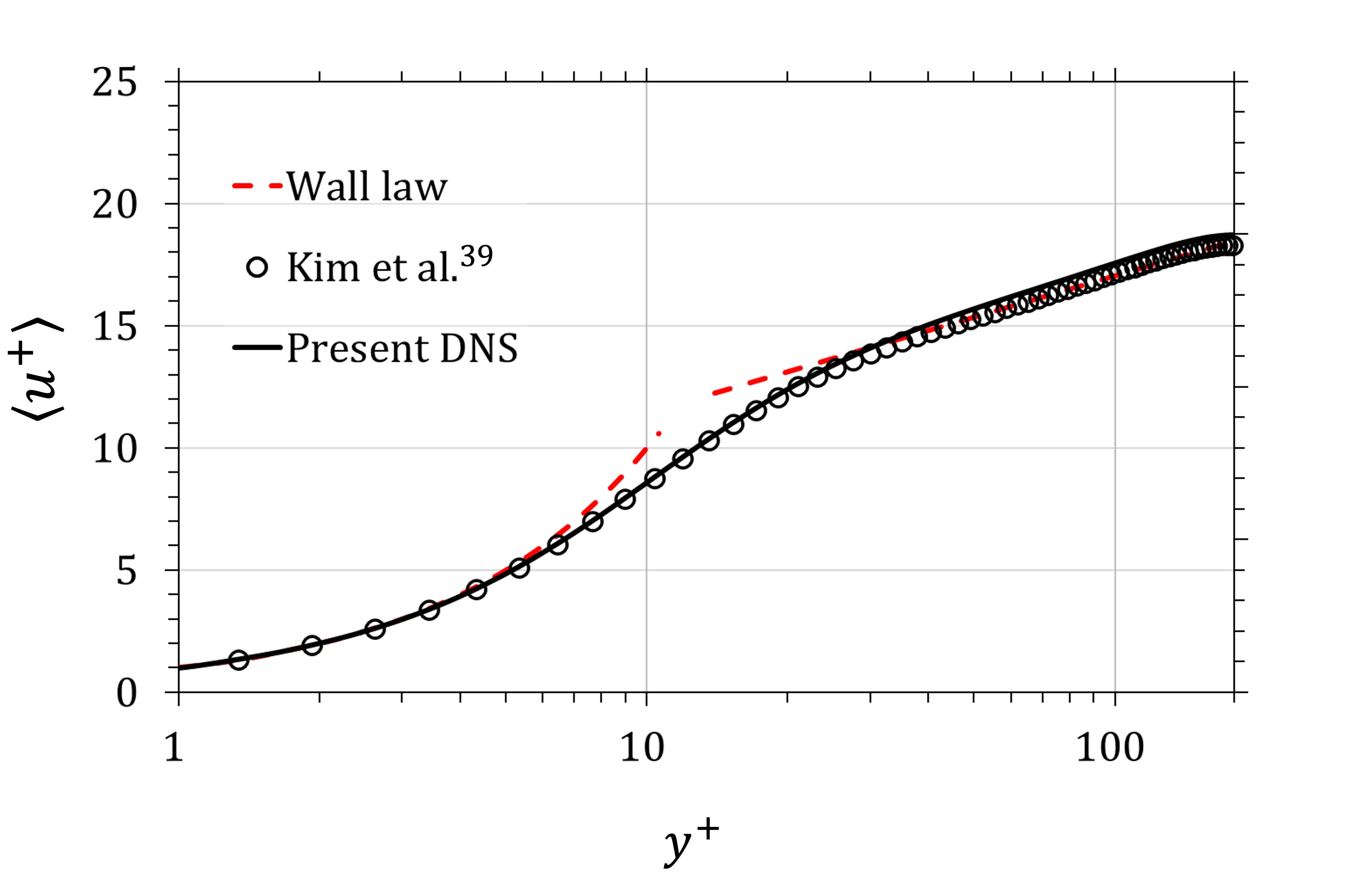}
            \caption{}
            \label{Fig2a}
        \end{subfigure}
        \hfill
        \begin{subfigure}{0.45\textwidth}
            \centering
            \includegraphics[width=\linewidth]{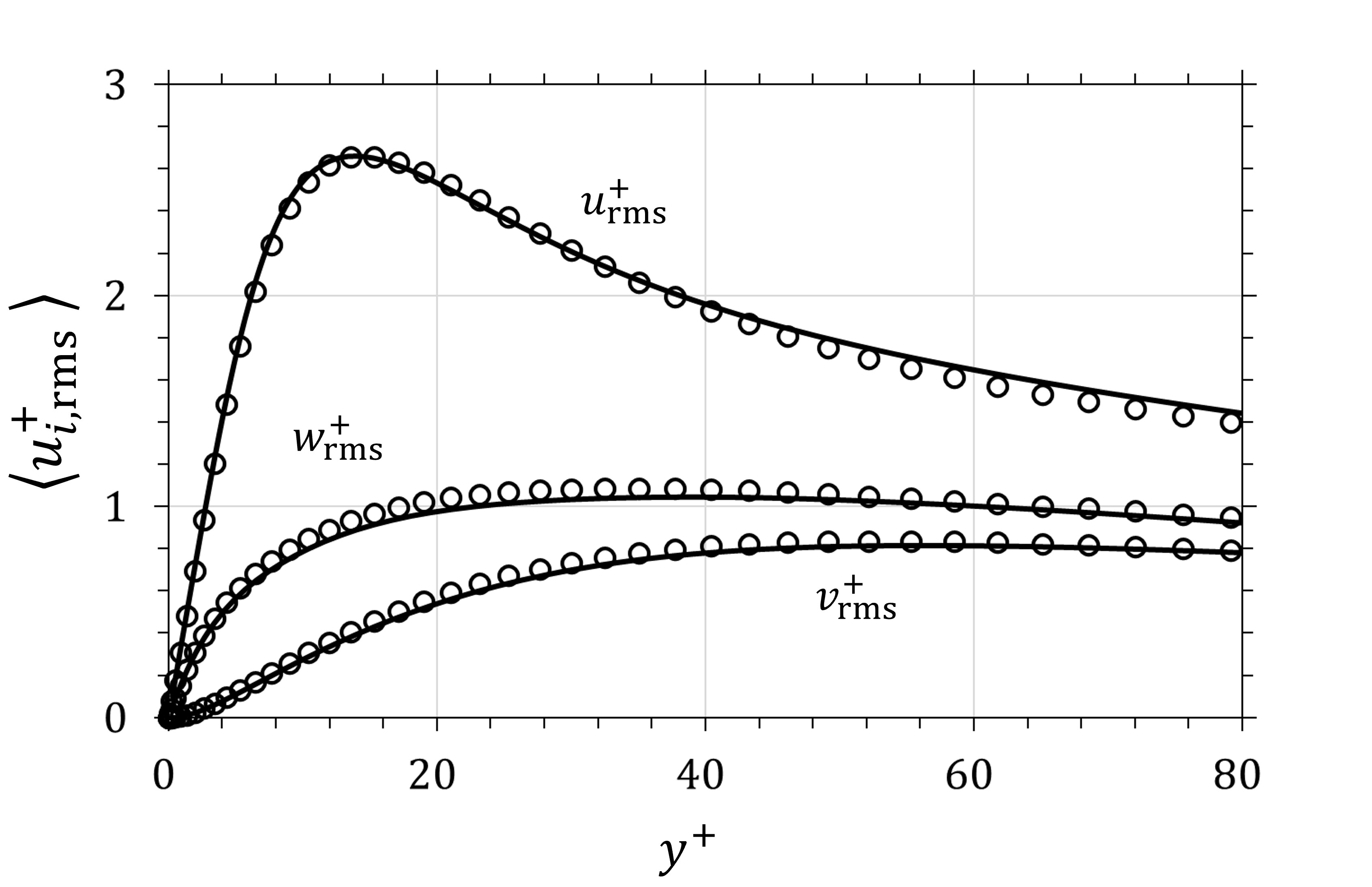}
            \caption{}
            \label{Fig2b}
        \end{subfigure}
        \caption{Wall-normal distribution of turbulence statistics of (a) mean velocity and (b) root-mean-square velocity. Here, $\langle \cdot \rangle$ denotes spatial averaging over the $x$--$z$ plane and temporal averaging; these notations are used hereinafter in this paper.}
        \label{Fig2}
    \end{figure*}

    The dataset for the data-driven SGS model is obtained from the DNS data of the turbulent channel flow with the friction Reynolds number $\text{Re}_{\tau} = \frac{u_{\tau} \delta}{\nu}=180$.  $u_{\tau}$ denotes the frictional mean velocity, $\nu$ is the kinematic viscosity, and $\delta$ is the channel half-width. The governing equations for incompressible continuity and Navier$-$Stokes equations are directly solved. A collocated grid with non-uniform spacing in the wall-normal ($y$) direction is used, whereby the no-slip boundary condition is applied to the walls and periodicity is applied in the streamwise ($x$) and spanwise ($z$) directions. The computational domains and computational grids are $(L_x \times L_y \times L_z) = (4\pi\delta \times 2\delta \times 2\pi\delta)$ and $(N_x \times N_y \times N_z) = (192 \times 128 \times 160)$, respectively. The unsteady solver employs a fractional-step method with second-order central differences for convective and viscous terms. The Adams$-$Bashforth method was adopted for time marching. The validation results (Fig.~\ref{Fig2}) show the wall-normal distributions of the mean streamwise velocity ($u^+$) and root-mean-square (rms) values of the fluctuating velocity components in the streamwise ($u_{\mathrm{rms}}^+$), wall-normal ($v_{\mathrm{rms}}^+$), and spanwise ($w_{\mathrm{rms}}^+$) directions, which are in good agreement with the results of Kim et al.~\cite{kim1987} and wall law. Here, the superscript $(.)^+$ denotes scaling in wall units.

    The DNS flow fields are then box-filtered to separate the GS and SGS quantities, yielding the fDNS data. The GS variables are used as the inputs, whereas the SGS stress, $\tau_{ij}$, is calculated to provide the corresponding label data. As in the study by Jalaali $\&$ Okabayashi\cite{jalaali2025}, the datasets are comprised of 140 instantaneous fields, approximately 1,820 non-dimensional time units ($tU_{bulk}/\delta$), 25,180 viscous time scales ($t u_{\tau}^2 / \nu$), and 291 flow throughs ($tU_{bulk} /L_x$), where $U_{bulk}$ denotes the bulk velocity. Here, the datasets are split 90:10 for training and testing.
    
    \subsection{\label{sec:level2} Baseline data-driven SGS model algorithm and framework}
\begin{figure}
    \includegraphics[width=1\linewidth]{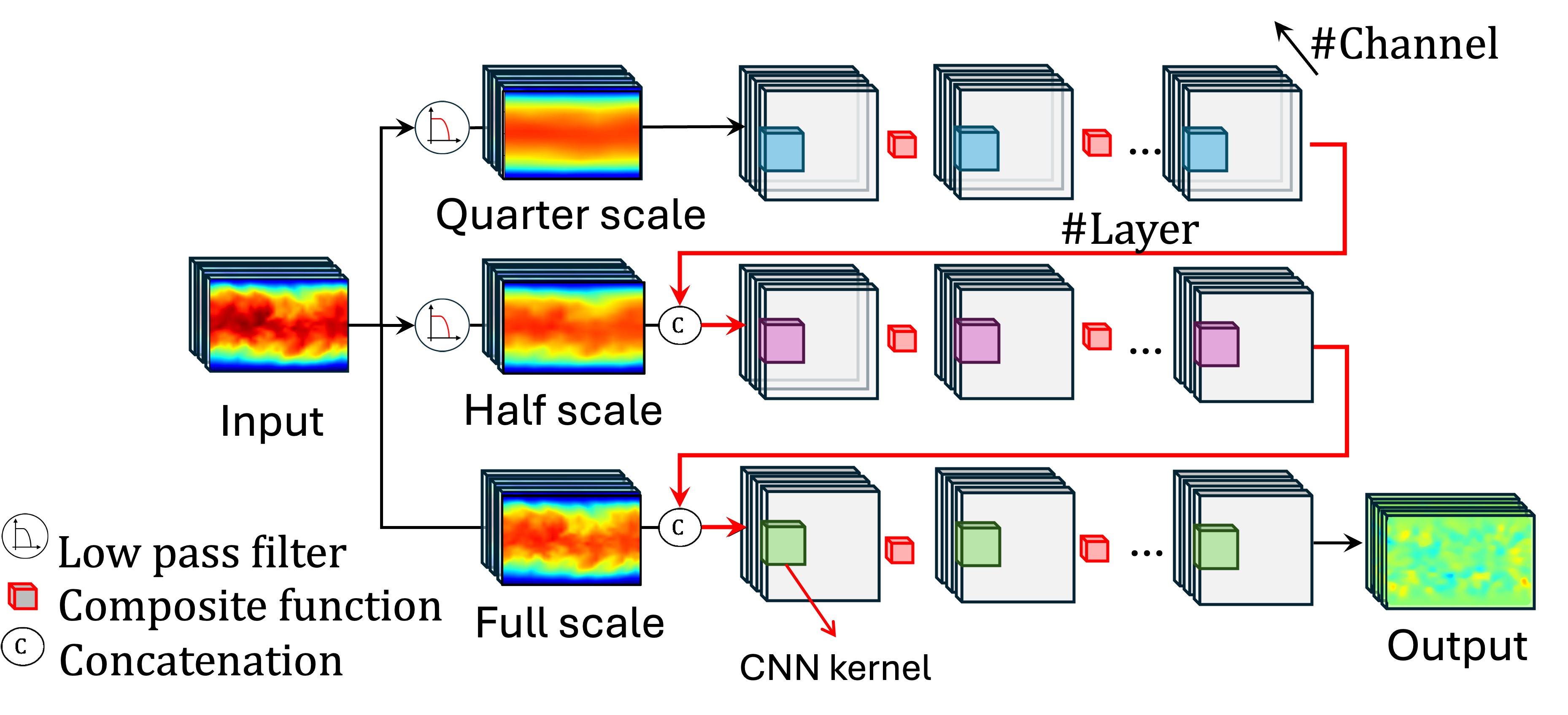}
    \caption{\label{Fig3} Schematic illustration of the DNN algorithm of the MSC model.}
    \end{figure}

    The DNN algorithm plays a crucial role owing to its ability to establish a functional relationship between the input features derived from the resolved flow fields and the six output variables $\tau_{ij}$. In this study, we adopted the multiscale-CNN model proposed by Jalaali and Okabayashi \cite{jalaali2025} with a minor change in the loss function (hereafter referred to as the MSC model) to represent multiple scales of a turbulent flow, which serves as the baseline architecture. We further developed two modified versions of this model, referred to as MSC2 and MSC3, in the subsequent sections to enhance rotational invariance. Jalaali $\&$ Okabayashi\cite{jalaali2025} emphasized that the MSC model incorporated multiscale representation and physical processes across scales, resulting in more accurate predictions. Here, the MSC model is outlined briefly and depicted in Fig.~\ref{Fig3}. Based on the principle of energy cascade, where energy is transferred from larger to smaller scales, the input features were decomposed into separate scales representing large-, intermediate-, and small-scale eddies. The MSC model encodes the information progressively, starting from the largest scale and proceeding to the smallest. Here, we found that the use of three scales provided the most accurate predictions, offering an optimal tradeoff between model complexity and accuracy. In each stage, the encoded information from the previous scale is concatenated to form a comprehensive multiscale representation. This concatenation is important for integrating the inter-layer information without sacrificing detail, as noted by Huang et al.\cite{huang2018}. 

    Based on the CNN algorithm, the convolutional layer computes the feature maps by convolving the input data with a kernel to detect the spatial patterns. The basic operation of a CNN is expressed as follows:
        \begin{equation}
    x_{k,l} =  \varphi \left(\text{BN}\left( w * {X}_{k,l-1} + b\right) \right).
    \label{eq6}
    \end{equation}
    where $w$ denotes the DNN weight or convolutional kernel in CNN; \text{BN} and $b$ denote batch normalization and bias, respectively. Here, $(.)_l$ denotes the hidden layer index $(l = 0 - 5)$ and  $(.)_k$ is the scale layer index; $k = 1,2$, and $3$ represent the quarter-, half-, and full-scale layers, respectively. Therefore, the output $\tau^p_{ij}$ is $x_{3,5}$. 3D convolutional kernels with a uniform size of 3 in each spatial direction are employed in the CNN to represent the nature of turbulent flow interactions. The exponential linear unit (ELU) activation function \( \varphi \) is applied, which introduces nonlinearity through a fixed mathematical mapping that transforms the data\cite{goodfellow2016}. ELU offers the advantage of a smooth gradient during backpropagation by allowing negative outputs, thereby helping to mitigate the vanishing gradient problem\cite{Clevert2015Fast,kim2020}. This property is particularly beneficial in capturing both the positive and negative components of the input variables. $\bar{D}_{ij}$ is chosen as the input variable, as it satisfies both Galilean and rotational invariance\cite{kajishima2017}. Hence, $X_{1,0}$ is $\bar{D}_{ij}$. As emphasized by various researchers, such as Spalart\cite{Spalart2023}, Brenner et al.~\cite{brener2024}, Wu et al.~\cite{wu1028}, and Prakash et al.~\cite{prakash2022}, the adherence of the input to the material objectivity requirement is critical for constructing data-driven SGS models
 
    \subsection{\label{sec:level2}Data-driven SGS model with rotational invariance}
    In the following, we describe the proposed model and its distinction from the baseline MSC model to satisfy the principle of material objectivity described in Eq.~\eqref{Eq1}. We first note that relying solely on invariant input variables is not sufficient to ensure material objectivity, because the DNN operates as a composite function, as formulated in Eq.~\eqref{eq6}. As noted by Mo $\&$ Zhao\cite{Mo} and Jaderberg et al.\cite{Jaderberg}, CNN models have limitations in handling rotational orientation, requiring modification of the model. Here, in contrast to the MSC model, both bias and BN terms were excluded, as follows:
        \begin{equation}
    x_{k,l} =  \varphi \left( w * {X}_{k,l-1}\right ).
    \label{eq8}
    \end{equation}
    This modification provides the data-driven SGS model with raw data representations, unaltered by normalization or bias, following the approach of Bin et al.\cite{Bin2022} and Cho et al. \cite{Cho2024} Physically, the inclusion of a bias term is unnecessary because the SGS stress output should vanish when the input tensor is zero\cite{Bin2022, Cho2024}. In addition, these studies showed that removing the bias and BN improved the generalization of the data-driven SGS model, which enabled the model to make predictions beyond the training data\cite{Bin2022, Cho2024}. Although BN is intended to improve the stability and convergence during training by independently normalizing the output of the CNN layers across mini-batches\cite{ioffe2015batch}, we argue that this partitioning of the mini-batches breaks the spatial coherence of the data and disturbs the underlying physical representation of the vortex interaction. It is important to note that such a modification may result in an unphysical representation, as the local interactions are correlated in turbulent flows. As a result, the data-driven SGS model becomes more fitted to the training dataset and may lose its ability to generalize under rotational transformations owing to the disrupted spatial correlation across the mini-batches. We also provide a separate discussion on the effects of removing bias and BN in Appendix~A. Therefore, by excluding the bias and BN terms while preserving the data structure through ELU activation, the convolutional kernel can be regarded as translationally invariant \cite{goodfellow2016}. However, rotation introduces a different type of transformation that is not automatically considered by the model. Accordingly, we hypothesize that rotation transformation takes place only at the initial input layer. As the input is $\bar{D}_{ij}$ and the rotational transformation is applied prior to the first convolutional layer, the effect of rotation is confined to the input stage. Because $\bar{D}_{ij}$ is an objective quantity, such transformations do not compromise the extracted features. Moreover, the ELU activation preserves both positive and negative values of the input, ensuring that no data are discarded and the underlying physical behavior of the features is not distorted. This newly proposed model configuration is hereinafter referred to as MSC2.
    
\begin{figure}
    \includegraphics[width=0.8\linewidth]{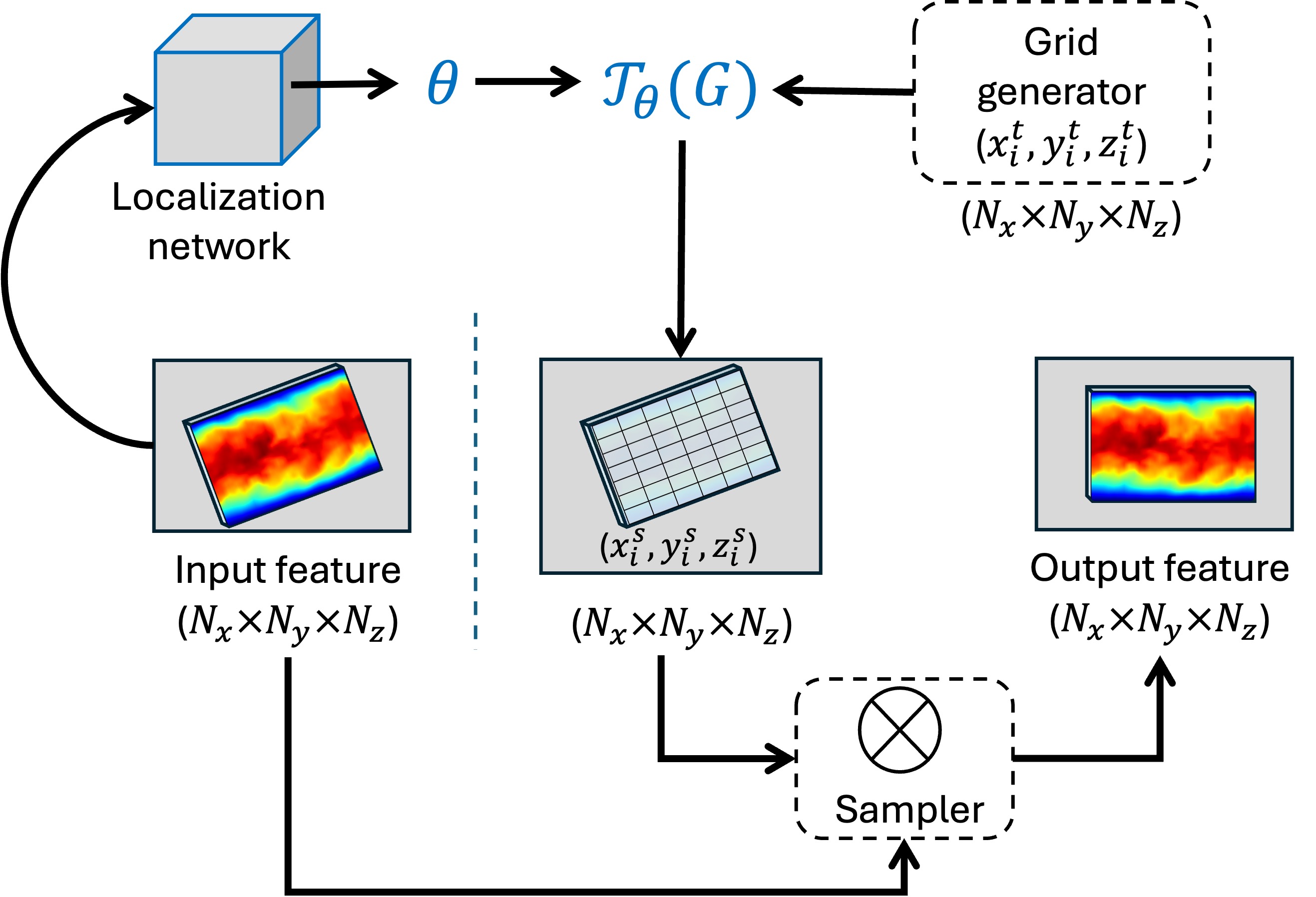}
    \caption{\label{Fig5} Schematic of STN algorithm based on Ref.\cite{Jaderberg}}
    \end{figure}

\begin{figure}
    \includegraphics[width=1\linewidth]{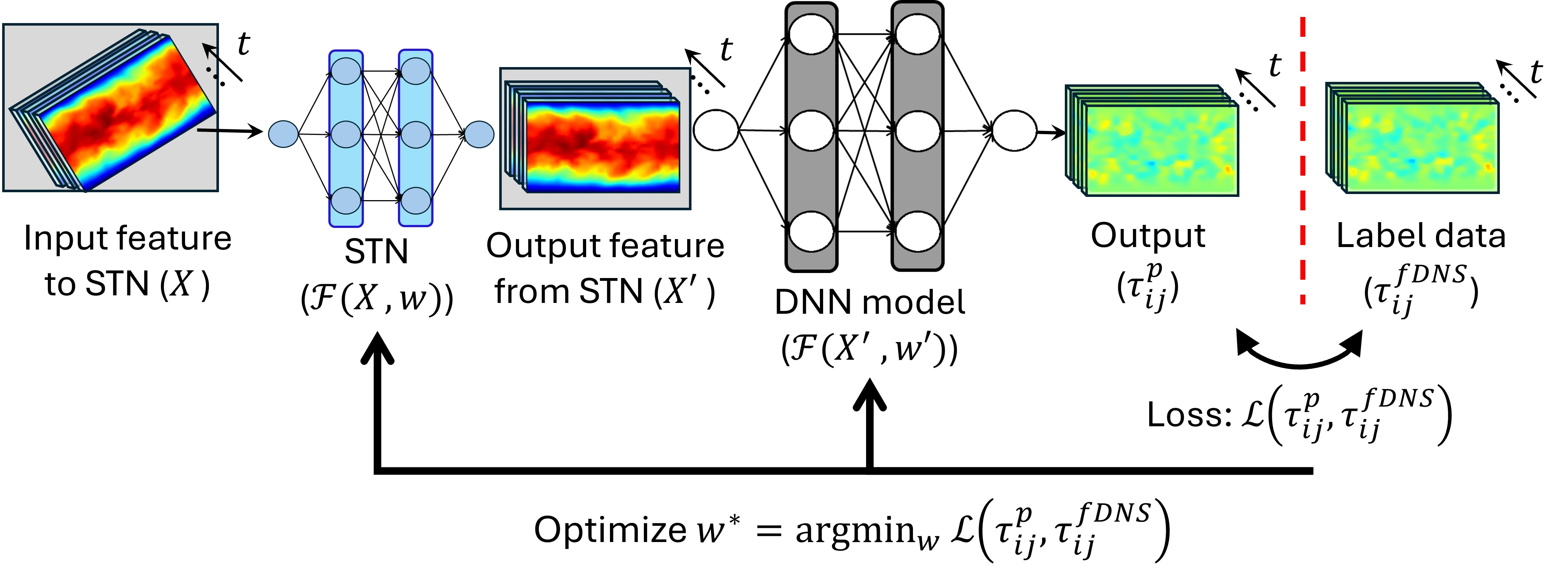}
    \caption{\label{msc_stn} Schematic of the MSC3 model.}
    \end{figure}

    To further enhance the invariance properties and accuracy of the data-driven SGS model, we incorporated an STN\cite{Jaderberg}. Specifically, the MSC2 model was modified by adding the STN algorithm (Fig.~\ref{Fig5}), hereinafter referred to as the MSC3 model (Fig.~\ref{msc_stn}). The main intention was to equip the data-driven SGS model with a transformation coordinate framework to help satisfy the principle of material objectivity. This approach is based on the assumption that, when given rotated inputs, a DNN algorithm that is sensitive to orientation may alter the spatial representation of the data regardless of the input quantity. The data-driven SGS model considerably lacks a spatial coordinate because neither the input nor the label data contain explicit coordinate information. Consequently, its ability to interpret the orientations of the flow structures is limited. Therefore, the STN algorithm is introduced to infer the geometric transformations and ensure spatial consistency. A concise description of the STN algorithm is provided here for completeness. A detailed explanation of its architecture and implementation can be found in the study by Jaderberg et al.\cite{Jaderberg}

    The STN algorithm consists of three main components: a localization network, grid generator, and sampler. The localization network first predicts the transformation parameters \((.)_\theta\) that describe geometric operations such as rotation or translation from the rotated and/or translated input feature. It is implemented as a regular CNN whose weights are optimized via backpropagation using the loss propagated from the data-driven SGS model (Eq.~\eqref{eq:loss1}), as illustrated in Fig.~\ref{msc_stn}. The loss function is minimized during training; a lower loss value indicates a more accurate prediction of the orientation, thereby allowing the network to infer the spatial transformation required to realign the input features. Through this process, the localization network learns to recognize the orientation of the input features. The predicted parameters \((.)_\theta\) are then passed to the grid generator, which provides a rectilinear grid of \( G = (x_i^t, y_i^t, z_i^t)\) corresponding to the \(i\)-th grid point that represents the grid coordinates in the target coordinate space of the STN output, where each grid point corresponds to a location to be sampled from the input features. Here, \( G\) consists of normalized coordinate values. The coordinate transformation between the output grid \( G\) and \((.)_\theta\) is performed using an affine transformation, denoted by \(\mathcal{T}_{(.)}\), which allows rotation and translation. The affine transformation for the 3D domain is defined as follows\cite{Jaderberg}:

    \begin{equation}
    \begin{pmatrix}
    x_i^{s} \\
    y_i^{s} \\
    z_i^{s}
    \end{pmatrix}
    = 
    \mathcal{T}_{\theta}(G_i)
    =
    \begin{bmatrix}
    \theta_{11} & \theta_{12} & \theta_{13} & \theta_{14} \\
    \theta_{21} & \theta_{22} & \theta_{23} & \theta_{24} \\
    \theta_{31} & \theta_{32} & \theta_{33} & \theta_{34}
    \end{bmatrix}
    \begin{pmatrix}
    x_i^{t} \\
    y_i^{t} \\
    z_i^{t} \\
    1
    \end{pmatrix}
    \label{STN}
    \end{equation}  

\begin{figure}
    \includegraphics[width=0.9\linewidth]{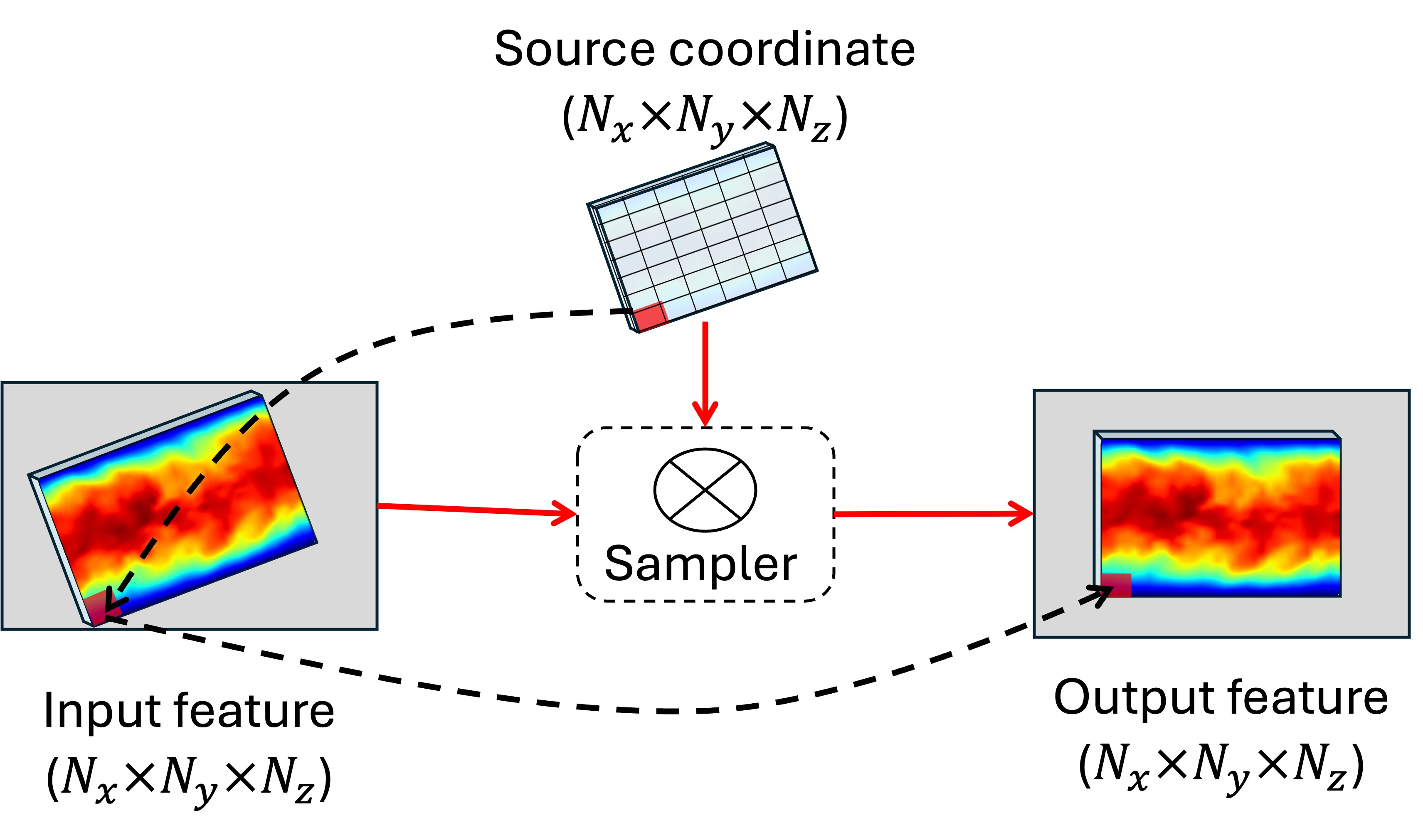}
    \caption{\label{sampler} Schematic of the sampler operator.}
    \end{figure}

    By applying the affine transformation to \( G\), the grid positions in the input feature are obtained as \((x_i^s, y_i^s, z_i^s)\), which denote the source coordinates corresponding to the \(i\)-th grid point. These coordinates represent the coordinate information of the given input feature. Because the source coordinates generally do not align with the discrete grid points on the output feature grid, a sampler is employed to estimate the feature values at those locations. The sampler performs trilinear interpolation over the neighboring voxels surrounding each source coordinate to compute the interpolated value corresponding to the output feature grid. As illustrated in Fig.~\ref{sampler}, the red arrows represent the flow of the sampler operation, whereas the black dashed line represents the mapping between the input and output features through the source coordinates. This ensures that the mapping between the input and output features remains spatially consistent \cite {Jaderberg}. As a result, the STN algorithm transforms the rotated and/or translated input features into spatially aligned (non-rotated) representations, allowing the data-driven SGS model to interpret the flow structures consistently across different orientations. This capability is particularly critical in turbulence modeling, where the accurate recognition of the spatial orientation and coherence directly affects the representation of the vortex interactions. This configuration is expected to enhance the invariance property of the data-driven SGS model. The model configurations and comparisons are summarized in Table~\ref{table:arch}.

    \begin{table*}
    \centering
    \small 
    \renewcommand{\arraystretch}{1.5}
    \caption{Comparison of data-driven SGS models.}
    \begin{tabular}{lccc}
        \toprule
        \text{Parameter} & MSC\cite{jalaali2025} & MSC2 & MSC3 \\
        \hline
        Exclusion of bias and BN & no & yes & yes \\
        STN & no & no & yes \\
        \toprule
    \end{tabular}
    \label{table:arch}
    \end{table*}

    \subsection{\label{sec:level2}Training of data-driven SGS model}
    During training, all data-driven SGS models were provided with the same dataset described in Sec.~II.B. However, for the MSC3 model, the training data were augmented by a ratio of 60:40 between the non-rotated and rotated datasets, whereas the total number of training data remained identical to that used for the MSC and MSC2 models. To train the localization network in the STN algorithm, this subset of the datasets was augmented by applying random rotations from $-30^\circ$ to $30^\circ$ around the spanwise direction. Note that only the input fields were rotated during the augmentation, whereas the corresponding label data were remained unrotated. The data-driven SGS model was optimized to determine the convolutional kernel weights $w$ that minimize the total loss function $\mathcal{L}$, defined as follows:

    \begin{equation}
    \begin{split}
    \mathcal{L} 
    &= \underbrace{\sum_{i=1}^{3} \sum_{j=1}^{3} 
    \gamma_{ij} \left(\tau_{ij}^{\text{fDNS}} - \tau_{ij}^{p} \right)^2}_{\mathcal{L}_d} \\
    &\quad + \underbrace{\sum_{i=1}^{3} \sum_{j=1}^{3} 
    \left( \langle \tau_{ij}^{\text{fDNS}} \rangle_{x,z} 
    - \langle \tau_{ij}^{p} \rangle_{x,z} \right)^2}_{\mathcal{L}_m} \quad  \\
    & + \underbrace{\frac{\lambda}{2} \sum_{k,l} w_{k,l}^{2}}_{\mathcal{L}_r}.
    \label{eq:loss1}
    \end{split}
    \end{equation}
The total loss function $\mathcal{L}$ consists of three components: the prediction loss ($\mathcal{L}_d$), mean loss ($\mathcal{L}_m$), and regularization loss ($\mathcal{L}_r$), as shown in Eq.~\eqref{eq:loss1}. Here, the loss function proposed by Jalaali $\&$ Okabayashi\cite{jalaali2025} for the MSC model is extended. An additional mean-loss term $\mathcal{L}_m$ is introduced to incorporate the wall-normal distribution information and improve the predictive accuracy in the near-wall region by following the approach of Arranz et al.\cite{arranz2024} Accordingly, $\gamma_{ij}$ is introduced to amplify the influence of each $\tau_{ij}$ component, thereby facilitating more effective updates to the kernel weights\cite{jalaali2025}. The values of $\gamma_{ij}$ are selected based on the global average magnitude of the respective $\tau_{ij}$ components, and further optimized through a hyperparameter tuning process. The selected values are $\gamma_{11} = 1$, $\gamma_{22} = 10$, $\gamma_{33} = 10$, $\gamma_{12} = 10$, $\gamma_{13} = 10$, and $\gamma_{23} = 100$. A regularization term $\mathcal{L}_r$ is included to prevent overfitting, with a regularization constant $\lambda$ set to $10^{-4}$. Here, $w_{k,l}$ denotes the convolutional kernel coefficients at scale $k$ and layer $l$.

To minimize the loss, we employ the adaptive moment estimation (ADAM) optimization algorithm~\cite{kingma2017}, which iteratively updates the convolutional kernel in the negative direction of the total loss gradient. Additionally, we apply a learning rate scheduler, with an initial learning rate of $10^{-3}$, which is reduced by a factor of 20 every 50 epochs. Here, an epoch refers to a complete pass through the entire training dataset. The data-driven SGS models were trained for 400 epochs with a mini-batch size of 8. The training was conducted using the PyTorch 1.13.1 machine learning library, and the computations were performed on a large-scale computer system consisting of eight NVIDIA A100 GPU cores, hosted at the D3 Center of the University of Osaka.
    
\section{Verification of material objectivity}
In this section, we discuss the predictive capability and physical consistency of the data-driven SGS model with rotational invariance using unseen data that differ from the training data. For this assessment, we rotated the input $\bar{D}_{ij}$ by $90^\circ$ and $270^\circ$ around the spanwise direction. Although such rotations do not correspond to any physical scenario in canonical channel flow, they serve as a robustness test for the data-driven SGS model under rotated-input conditions. Here, the evaluation of the data-driven SGS model is focused on the objectivity condition given by Eq.~\eqref{Eq1}, thereby for example, under a $90^\circ$ rotation, this relation is approximated as

\begin{equation}
    \mathcal{R}_{90} \left( \mathcal{F}(\bar{D}_{ij}) \right) - \mathcal{F}(\mathcal{R}_{90} \bar{D}_{ij}) \approx 0.
\label{equi2}
\end{equation}
where $\mathcal{R}$ denotes the rotation operator. It is important to note that, because the data-driven SGS model is constructed using a DNN algorithm, which is not a strictly mathematical function but rather a learned approximation, some deviations in Eq.~\eqref{equi2} are expected.

\begin{figure*}
    \centering
    \subfloat[\label{tau12_0}]{\includegraphics[width=0.4\linewidth]{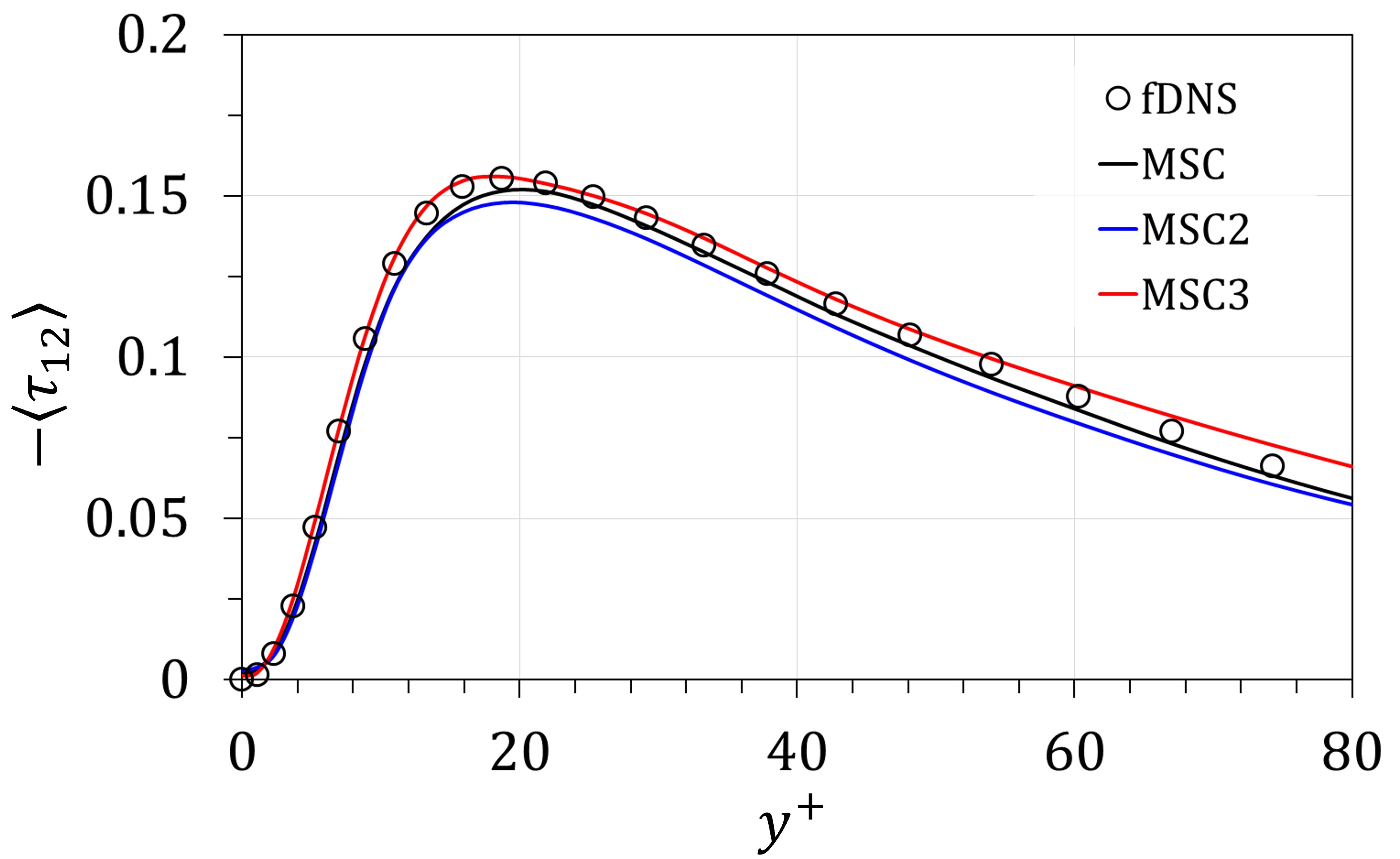}}\hspace{0.01\linewidth}
    \subfloat[\label{tau12_90}]{\includegraphics[width=0.4\linewidth]{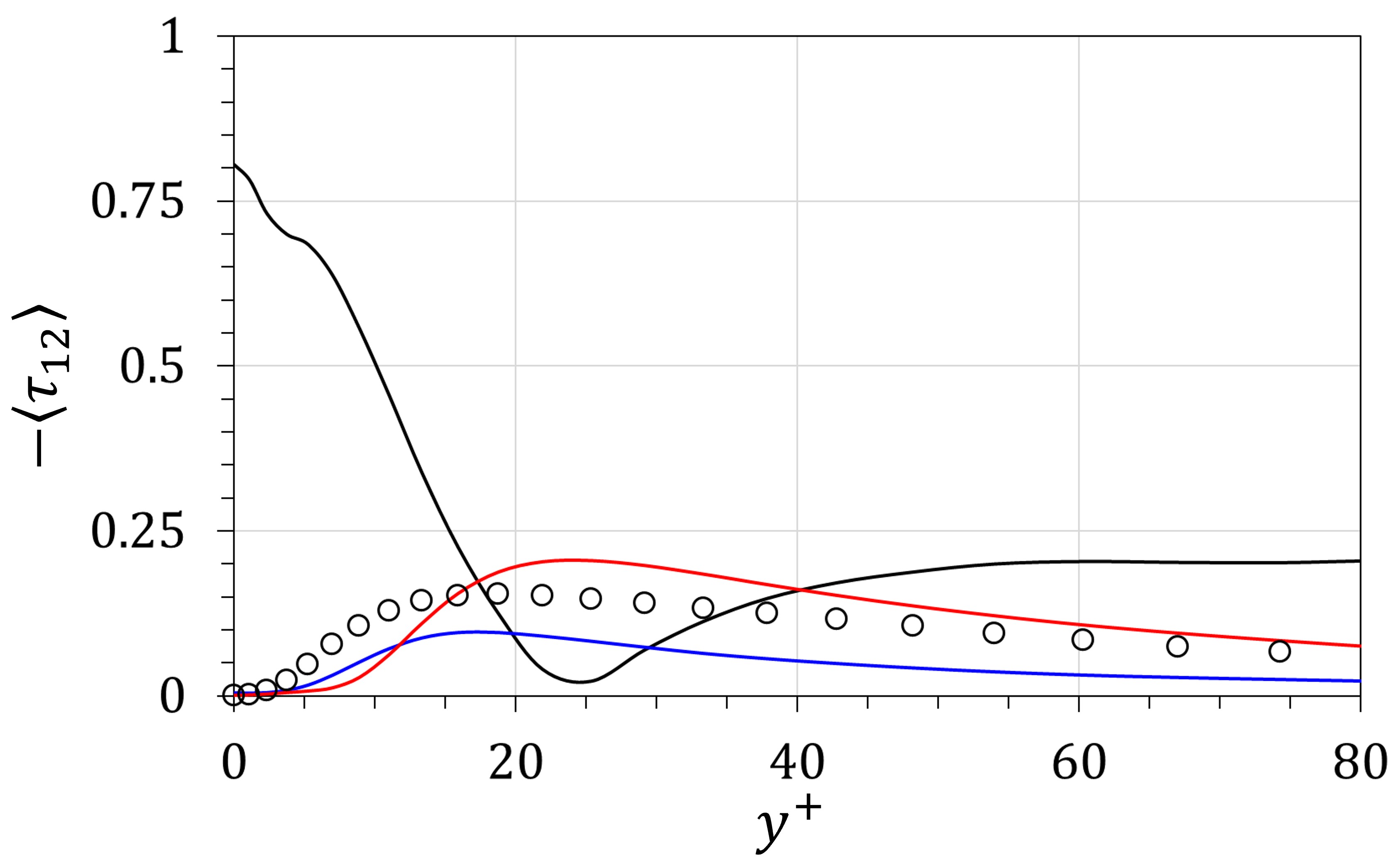}}\hspace{0.01\linewidth}
    \subfloat[\label{tau12_270}]{\includegraphics[width=0.4\linewidth]{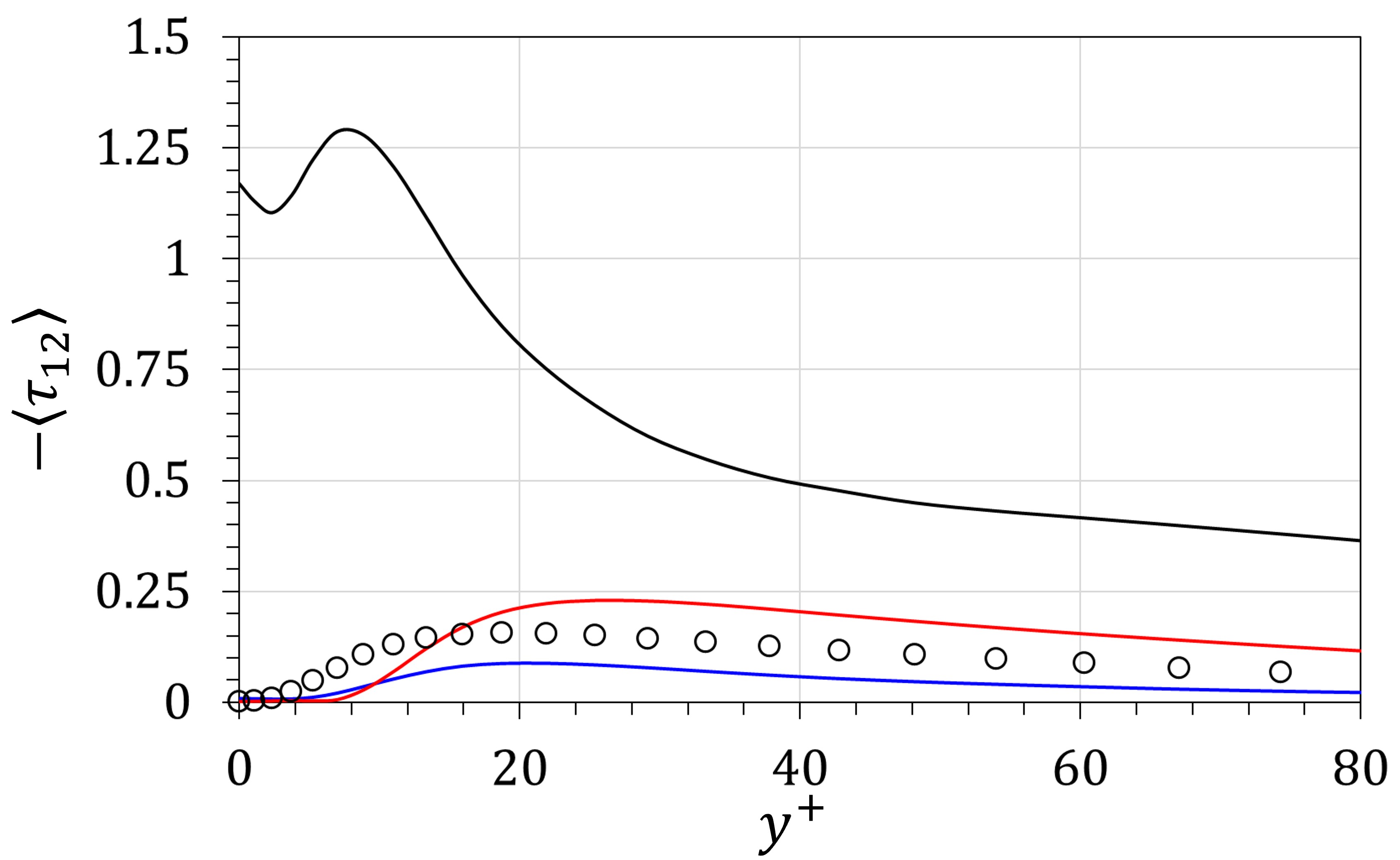}}
    
    \caption{Wall-normal distributions of $\tau_{12}$ with input rotated by (a) $0^\circ$, (b) $90^\circ$, and (c) $270^\circ$.}
    \label{Fig6}
\end{figure*}

First, the evaluation focused on $\tau_{12}$, which is considered the most significant component in wall-bounded turbulent flows~\cite{bose2024}. The wall-normal distribution of the shear stress component $\tau_{12}$ is depicted in Fig.~\ref{Fig6}. Here, the non-rotated input implies $0^\circ$ rotation, and the same expression is used hereinafter. For the non-rotated input, the overall agreement between the MSC model and fDNS data is comparable to that between the MSC2 and MSC3 models, although the MSC3 model exhibits a better prediction capability in the near-wall regions. In the region of $\textit{y}^+>20$, the MSC model can reasonably capture the overall trend of the fDNS distribution, whereas the MSC2 and MSC3 models exhibit a slight underprediction and overprediction, respectively. As also discussed in Appendix~A, this is likely attributed to the presence of bias and BN, allowing data-driven SGS model to fit the data distribution better. Despite the exclusion of both bias and BN, the results of the MSC2 and MSC3 models are comparable to those of fDNS. This indicates that removing bias and BN does not inherently degrade the model accuracy. For the rotated input, the MSC model shows significant deviations and is unable to predict accurately, which demonstrates that this model does not satisfy the requirement of material objectivity. In contrast, both the MSC2 and MSC3 models maintain consistent accuracy under rotation and show a slight deviation from the fDNS data. This result implies that the MSC2 and MSC3 models satisfy the requirement of objectivity or rotational invariance. 

Moreover, MSC3 achieves a closer alignment with fDNS than MSC2, which provides further support for the earlier assumption that rotation applied directly to the input tensor $\bar{D}_{ij}$ can distort the spatial relationships among its components, as DNN algorithms are not inherently mathematical functions (Eq.~\eqref{equi2}). Consequently, such a rotation compromises the prediction accuracy of the model, whereas introducing the STN algorithm helps improve the accuracy by preserving the canonical orientation of $\bar{D}_{ij}$. Moreover, this interpretation is further supported by the visual assessments presented in Fig.~\ref{Fig7}. The term \textit{err} is defined as the difference between the rotation of the data-driven SGS output and data-driven SGS output of the rotated input, similar to Eq.~\eqref{equi2}. The MSC2 and MSC3 models produced comparable predictions before and after the rotation, with error maps revealing relatively small discrepancies across the domain. To quantify the discrepancies, we computed the mean absolute error (MAE) over the entire domain. The MAE values for MSC, MSC2, and MSC3 are 0.438, 0.038, and 0.028, respectively. These results confirm that enforcing rotational invariance not only through the input features but also within the DNN architecture substantially improves the rotational invariance and physical consistency of data-driven SGS models.

    \begin{figure}
    \includegraphics[width=1\linewidth]{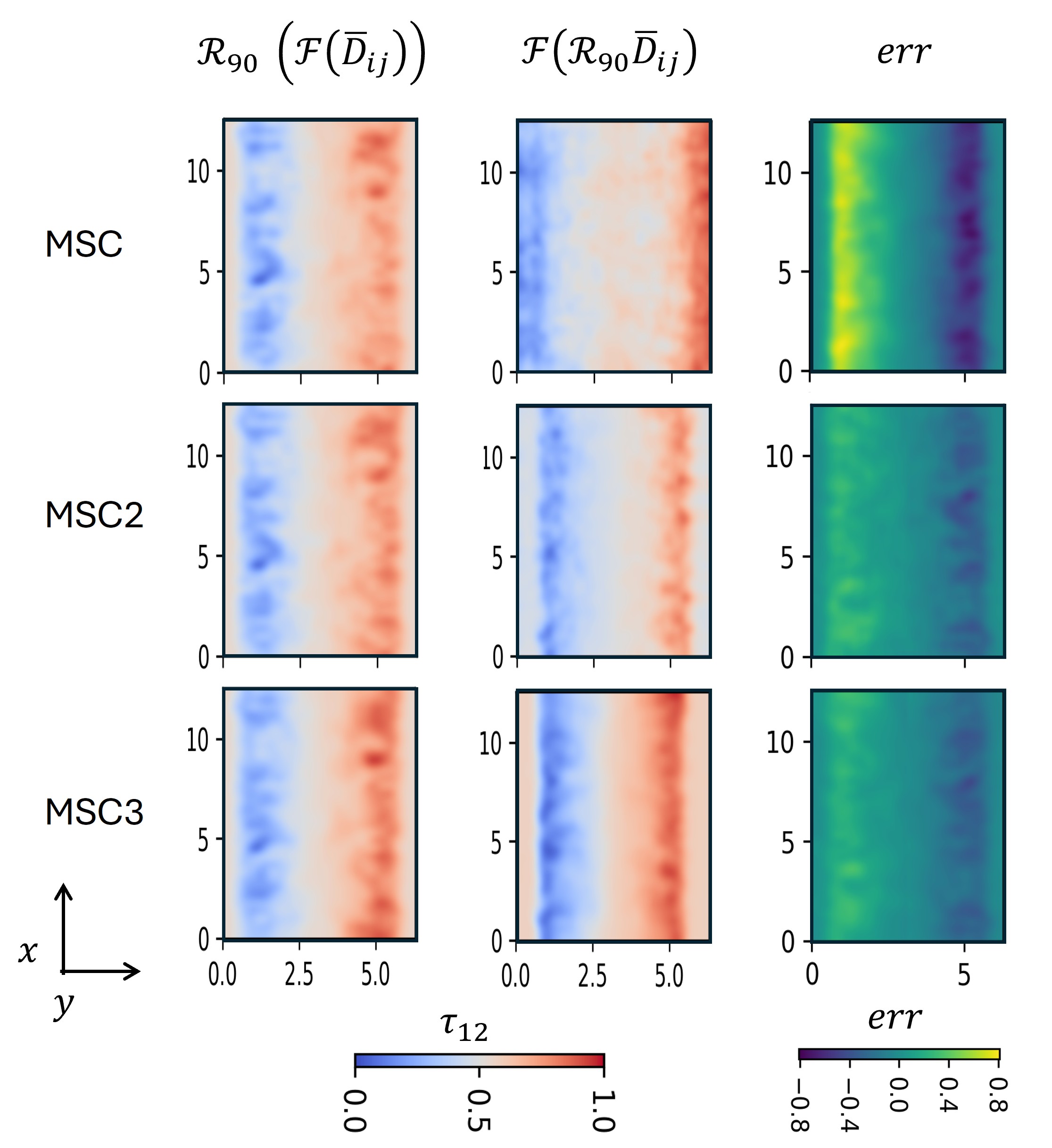}
    \caption{\label{Fig7} Instantaneous field of $\tau_{\text{12}}$ in \textit{x}$-$\textit{y} plane at \textit{z}$=L_z/2$ from \textit{a priori} test of the rotational invariance of the data-driven SGS model. The input is rotated by $90^\circ$.}
    \label{Fig7}
    \end{figure}

\begin{figure*}
    \centering
    \subfloat[\label{tau11}]{\includegraphics[width=0.4\linewidth]{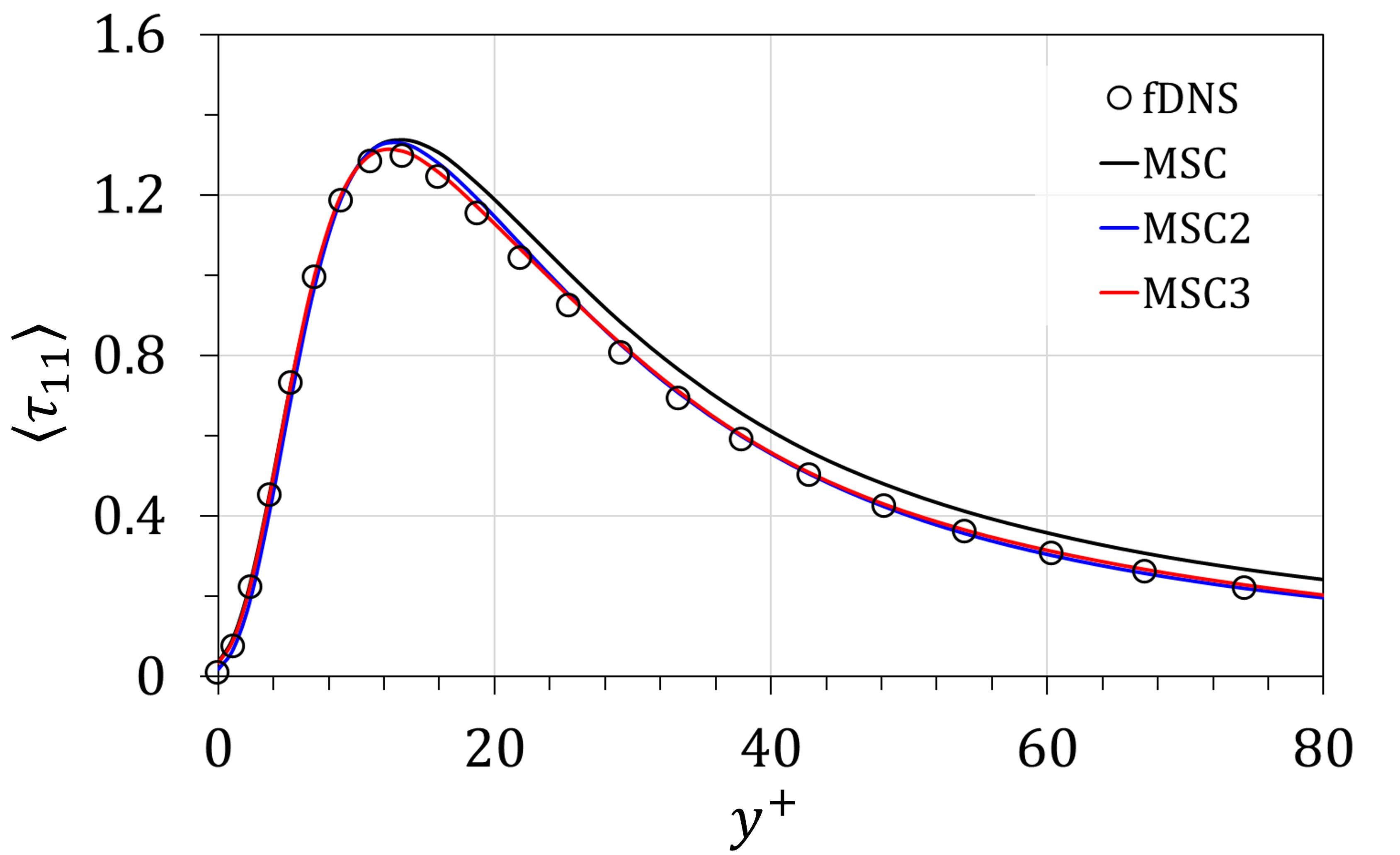}}\hspace{0.01\linewidth}
    \subfloat[\label{tau11_90}]{\includegraphics[width=0.4\linewidth]{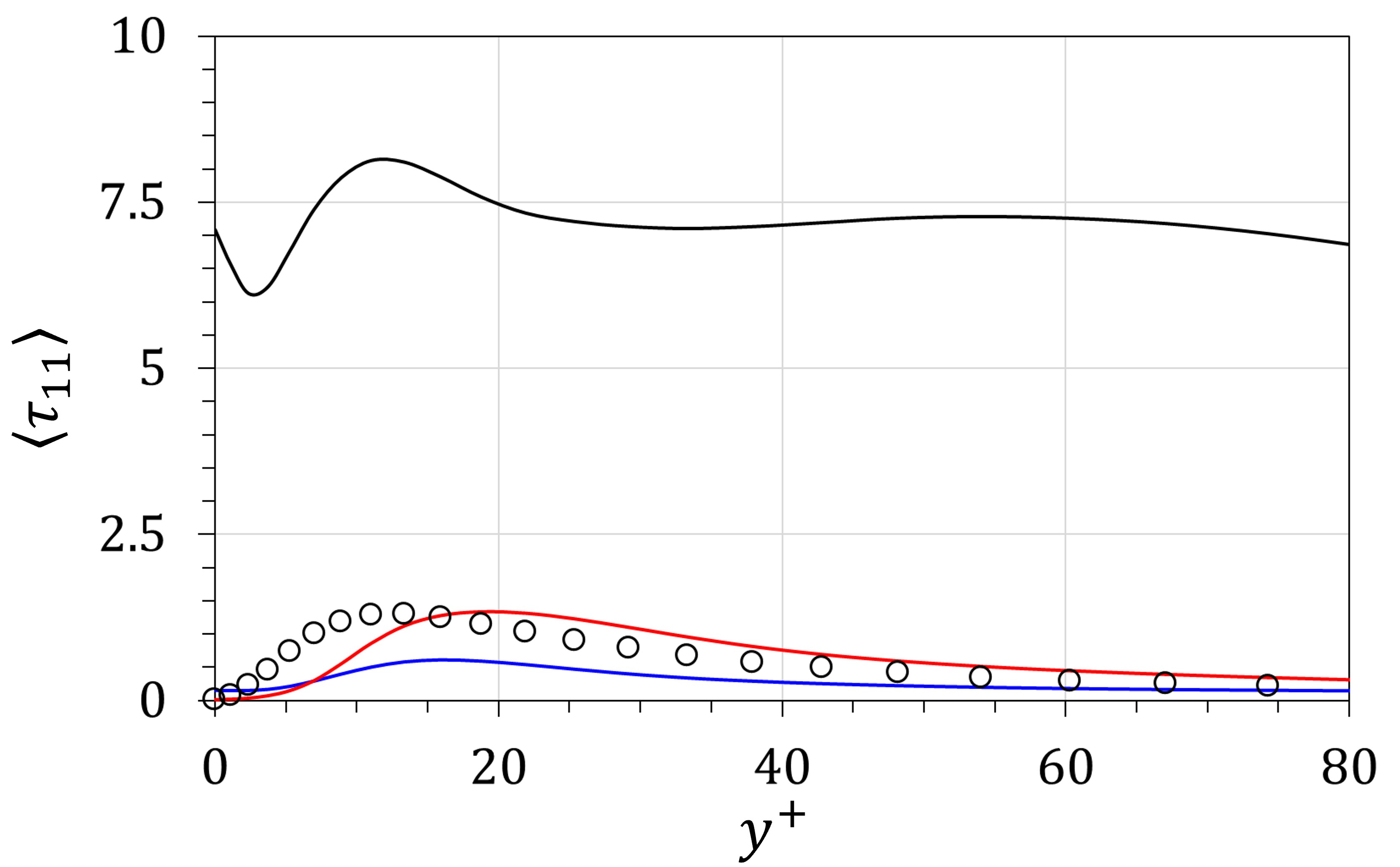}}\hspace{0.01\linewidth}
    \subfloat[\label{tau11_270}]{\includegraphics[width=0.4\linewidth]{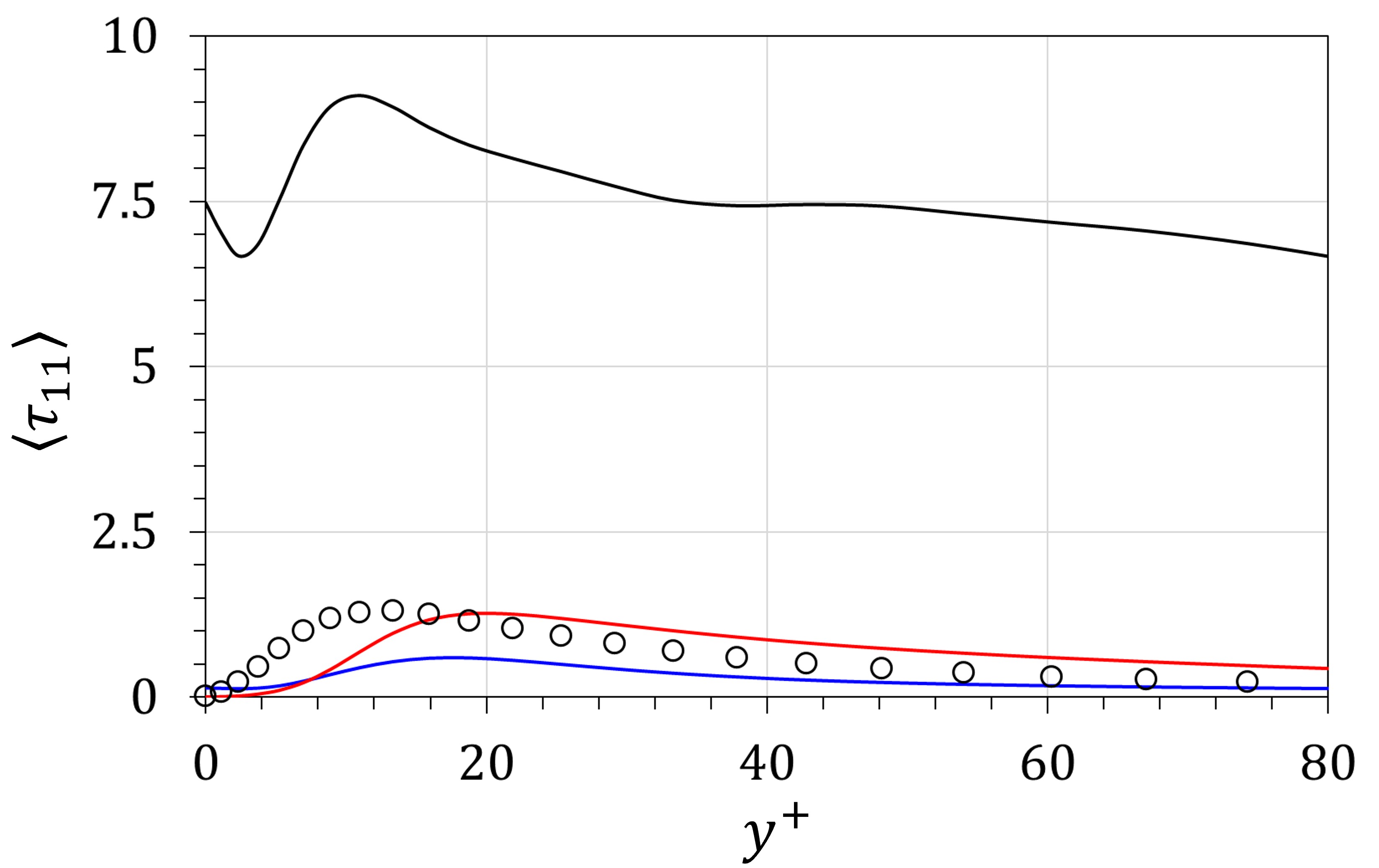}}
    
    \caption{Wall-normal distributions of $\tau_{11}$ with input rotated by (a) $0^\circ$, (b) $90^\circ$, and (c) $270^\circ$.}
    \label{tau11}
\end{figure*}

Furthermore, we evaluated the prediction performance on $\tau_{11}$, which quantifies the unresolved stresses associated with the streamwise velocity fluctuations. $\tau_{11}$ represents the streamwise component of the SGS stress, reflecting the energy transfer between the resolved and subgrid-scale kinetic energy in the main flow direction\cite{sagaut2006}. Following the same procedure as in the previous analysis, we assessed the performance of the data-driven SGS model for both non-rotated and rotated inputs. Fig.~\ref{tau11} shows the wall-normal distributions of $\tau_{11}$ for both non-rotated and rotated inputs. For the non-rotated input, all models successfully capture the distribution of $\tau_{11}$, showing excellent agreement with the fDNS data. In contrast, when the input fields are rotated, the MSC model fails to predict $\tau_{11}$ accurately. This discrepancy can be attributed to the sensitivity of $\tau_{11}$ to the input orientation. Unlike $\tau_{12}$, which represents a shear component between the upper and lower walls, $\tau_{11}$ might depend explicitly on the streamwise direction. Consequently, when the input is rotated, the streamwise direction changes its alignment relative to the learned coordinate system, making the prediction task significantly more difficult. Another possible factor is that $\tau_{11}$ exhibits a large value of the SGS stress, such that the rotation of the input field can lead to a more pronounced alteration in the corresponding feature representation. Nevertheless, both the MSC2 and MSC3 models are able to provide accurate predictions for the rotated input.

\begin{figure*}
    \centering
    \subfloat[\label{Dist11}]{\includegraphics[width=0.3\linewidth]{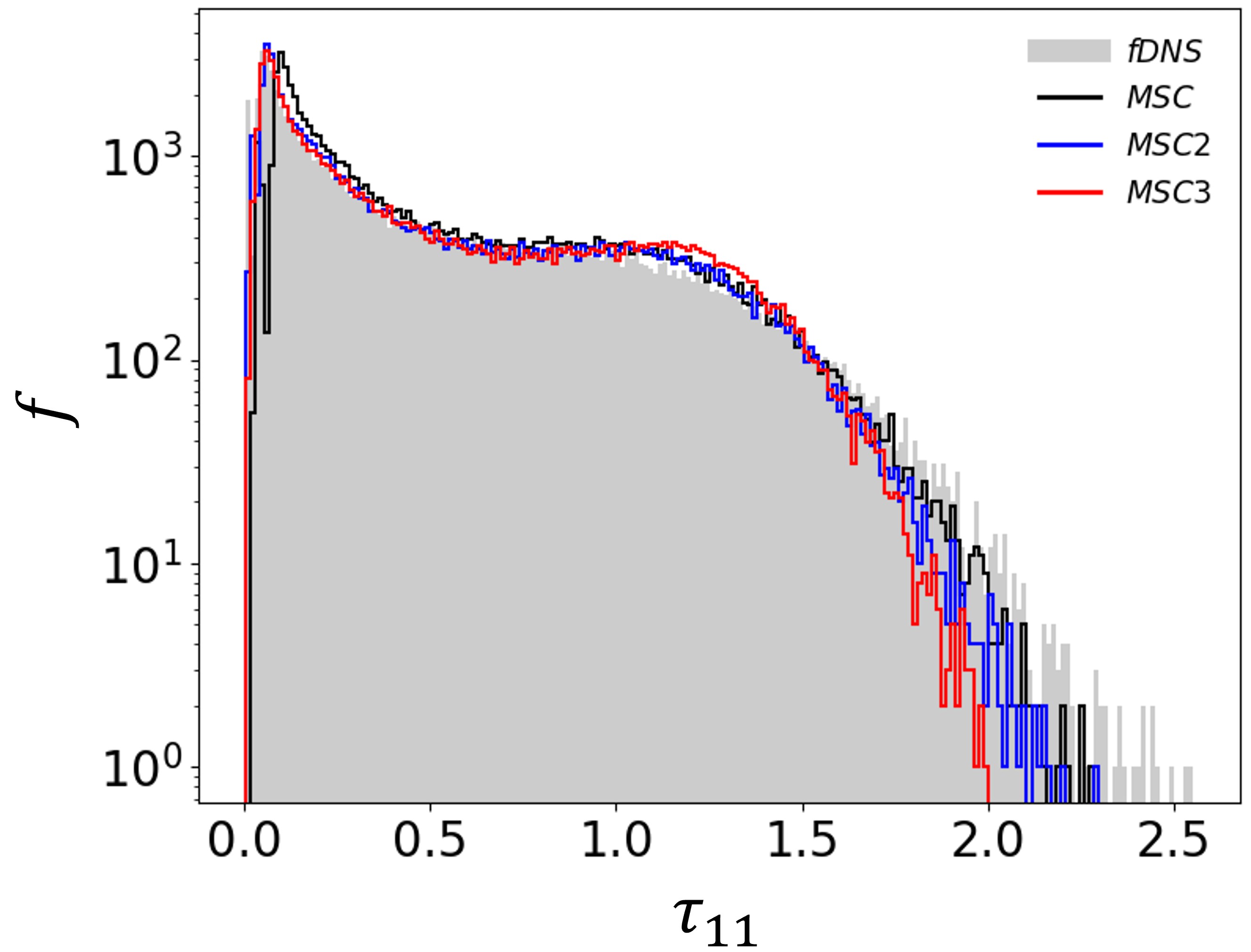}}\hspace{0.01\linewidth}
    \subfloat[\label{Dist1190}]{\includegraphics[width=0.3\linewidth]{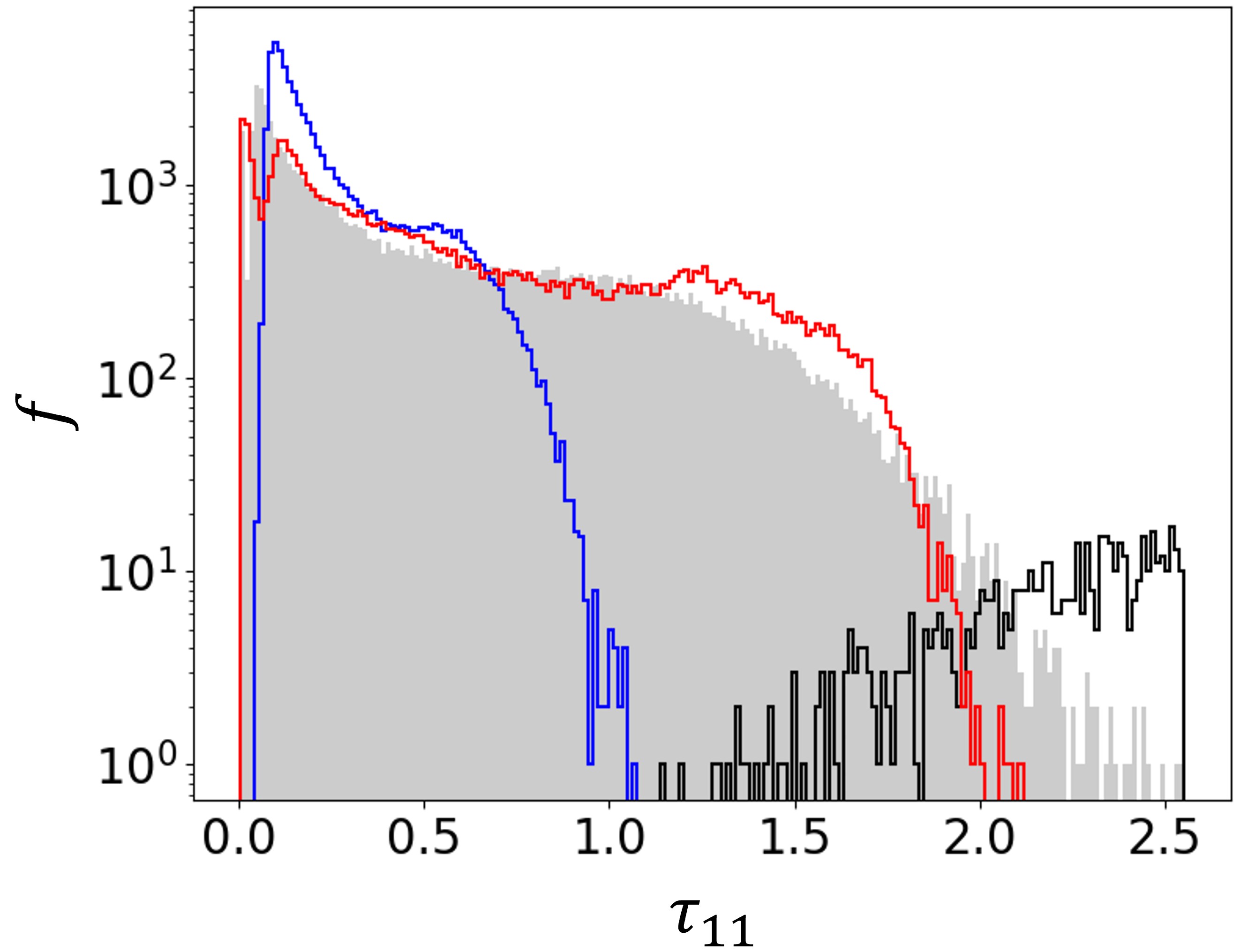}}\hspace{0.01\linewidth}
    \subfloat[\label{Dist11270}]{\includegraphics[width=0.3\linewidth]{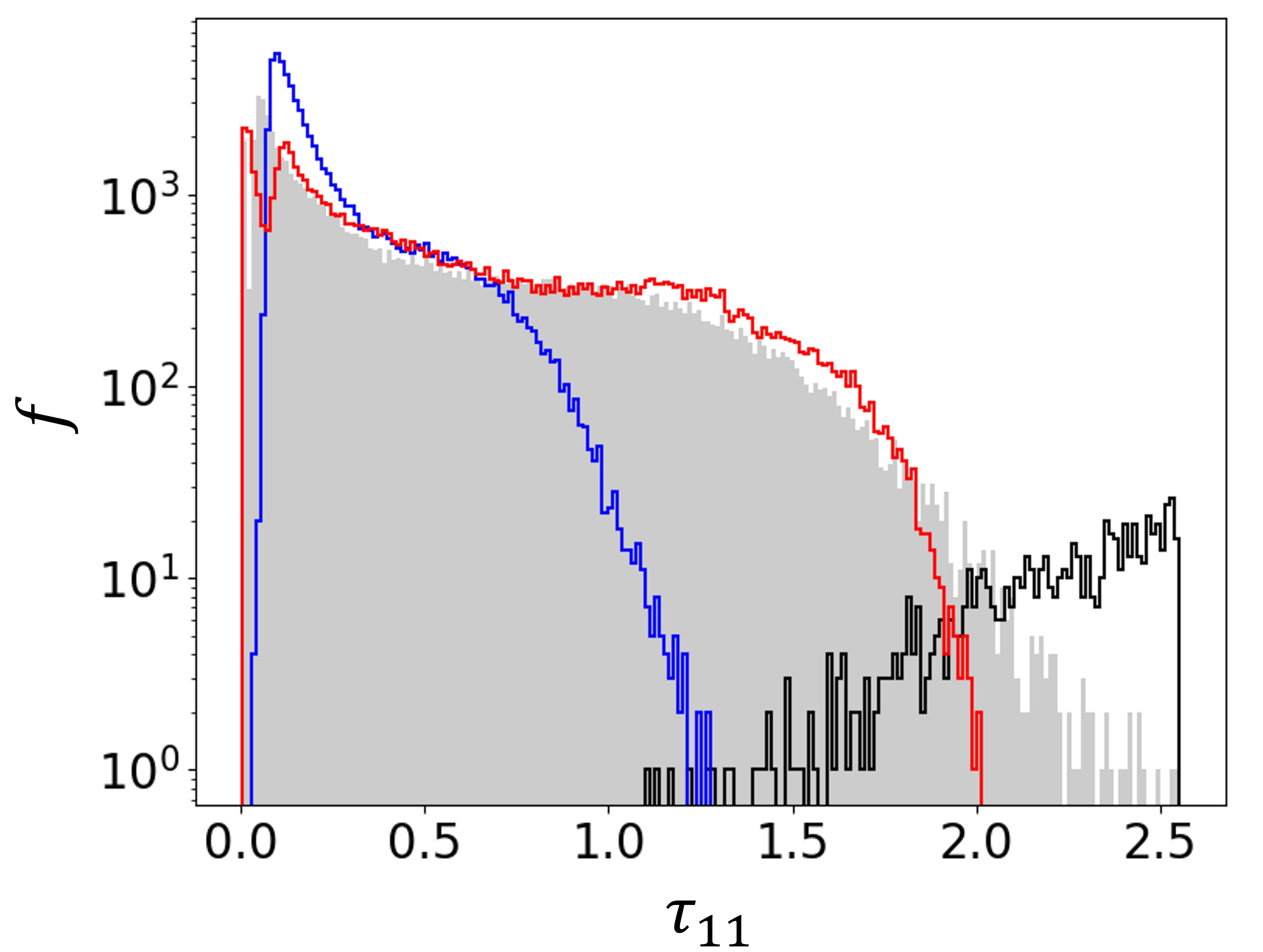}}\hspace{0.01\linewidth}
    \subfloat[\label{Dist12}]{\includegraphics[width=0.3\linewidth]{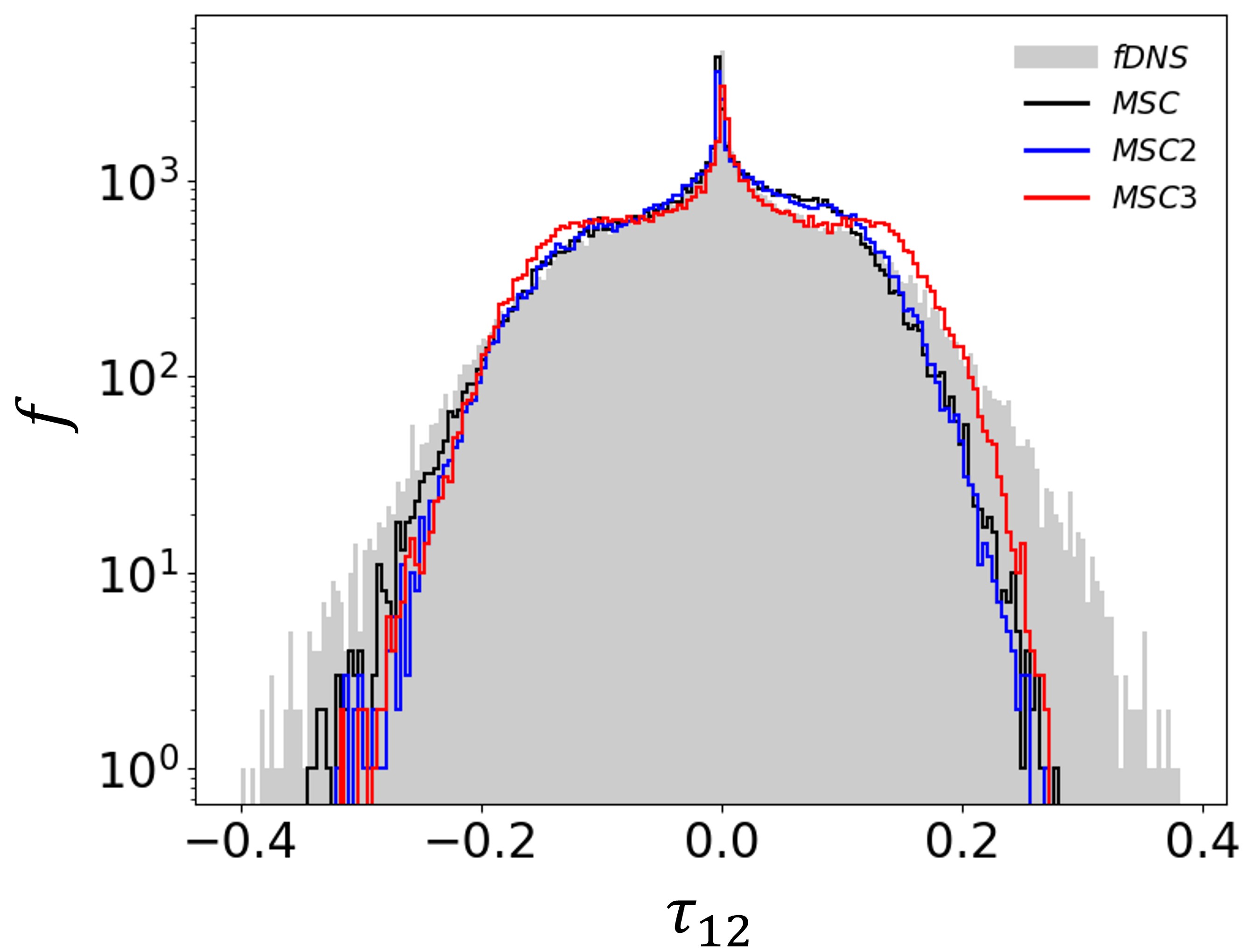}}\hspace{0.01\linewidth}
    \subfloat[\label{Dist1290}]{\includegraphics[width=0.3\linewidth]{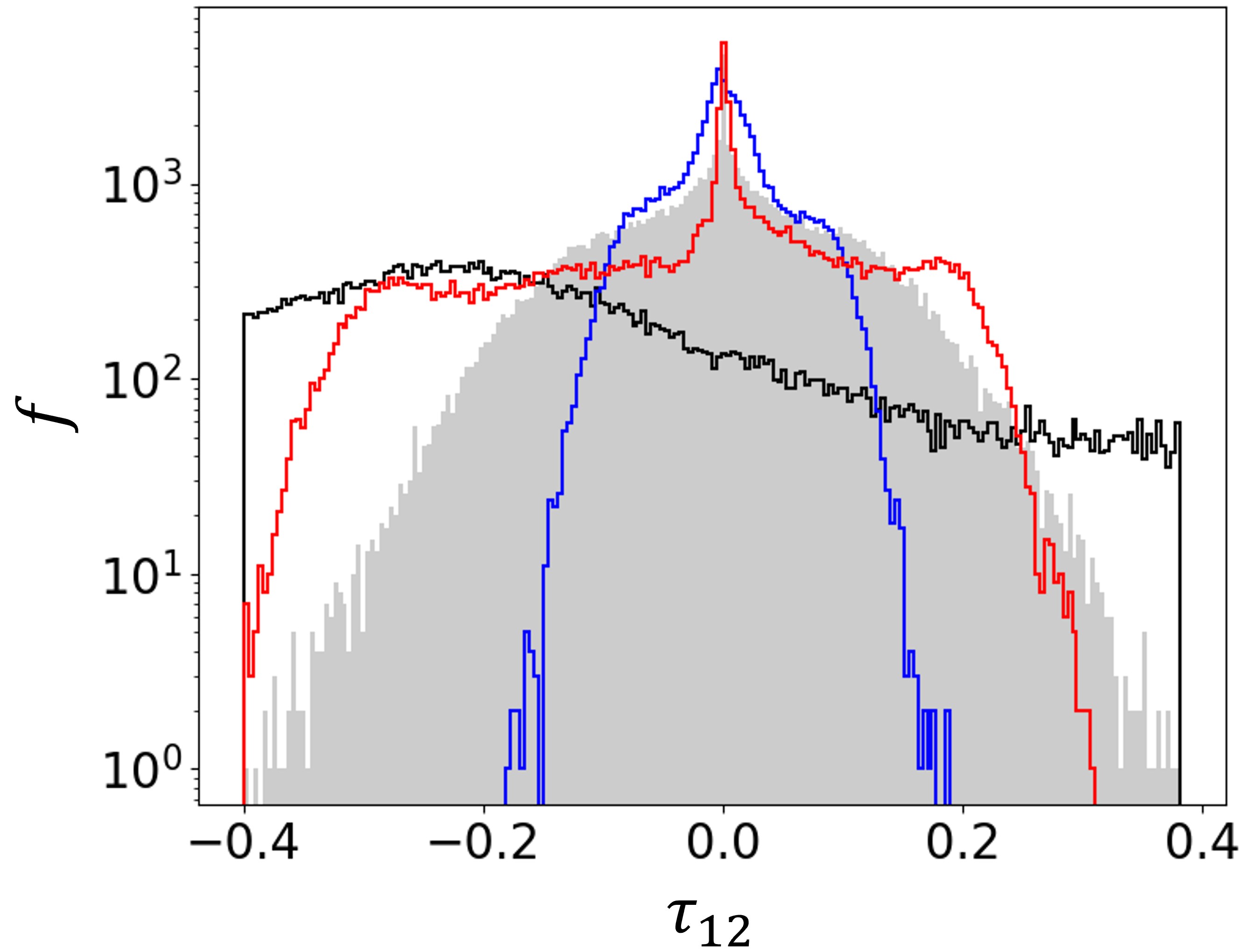}}\hspace{0.01\linewidth}
    \subfloat[\label{Dist12270}]{\includegraphics[width=0.3\linewidth]{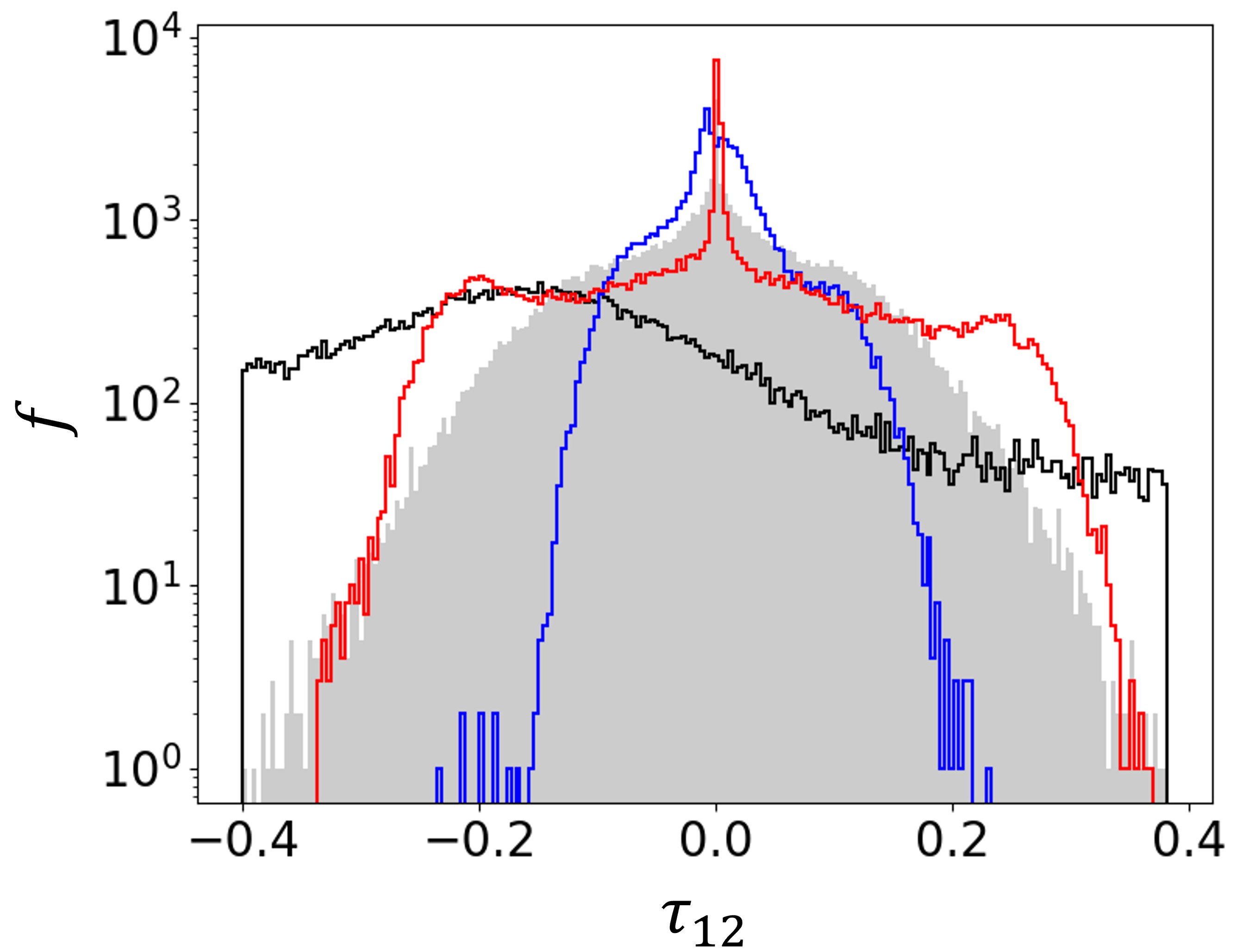}}
    
    \caption{Distribution of the number of occurrences $f$ of (a–c) $\tau_{11}$ and (d–f) $\tau_{12}$ with input rotated by (a, d) $0^{\circ}$, (b, e) $90^{\circ}$, and (c, f) $270^{\circ}$. The shaded regions denote the fDNS data.}
    \label{bins}
\end{figure*}

Figure~\ref{bins} presents the distribution of the number of occurrences $f$ for $\tau_{11}$ (top row) and $\tau_{12}$ (bottom row) at different rotation angles of $0^{\circ}$, $90^{\circ}$, and $270^{\circ}$. Here, the number of bins in the histogram is set to 200, which corresponds to the bin width of $0.01$. The $\tau_{11}$ and $\tau_{12}$ data within each bin are counted. $\tau_{11}$ shows a positive distribution with a pronounced right tail that indicates strong fluctuations. This positive contribution also significantly enhances the SGS turbulent kinetic energy\cite{sagaut2006}. The accurate prediction of $\tau_{11}$ can be crucial for maintaining the net forward energy transfer and SGS kinetic energy. The shear component $\tau_{12}$ takes both positive and negative values, reflecting the turbulent shear induced by the wall-normal velocity gradients, which is important in the momentum exchange of a turbulent flow\cite{davidson2015}. Underprediction of the value weakens the subgrid stress, whereas exaggeration of the tails results in overprediction of the shear.

For the non-rotated input, all models successfully reproduce both $\tau_{11}$ and $\tau_{12}$, indicating that the data-driven SGS models can capture the correct range of the subgrid stress magnitudes, despite an underprediction of both tails of the $\tau_{12}$ distribution. However, consistent with the wall-normal distributions in Figs.~\ref{Fig6} and \ref{tau11}, significant differences emerge when the input fields are rotated. The MSC model exhibits noticeable deviations from the fDNS data, particularly for $\tau_{11}$, where the predicted distribution shifts toward lower values and loses the characteristic positive tail observed in the fDNS results. As previously discussed, the degradation in $\tau_{11}$ prediction can be attributed to its strong dependence on the input orientation. When the coordinate frame is rotated, the streamwise direction is no longer aligned with the trained reference frame, leading to a mismatch between the learned mapping of the input features and the corresponding stress component. For $\tau_{12}$, a noticeable discrepancy also appears in the MSC model, which fails to reproduce the fDNS distribution. In addition, the MSC2 model shows a tendency to underpredict the number of occurrences of $\tau_{11}$ and $\tau_{12}$, suggesting a suppression of the SGS effects. This may lead to reduced turbulence interactions and weakened energy transfer between the resolved and subgrid-scales, which likely results in an underestimation of SGS dissipation. A prominent improvement is shown by the MSC3 model, which reproduces both $\tau_{11}$ and $\tau_{12}$ more accurately for the rotated input. However, for $\tau_{12}$, it shows discrepancies at both tails, where the negative tail is strongly overpredicted at $90^{\circ}$ rotation. This may be attributed to limitations in the STN algorithm, particularly its reduced ability to generalize the rotated input, possibly owing to limited exposure to the rotated input during training. Such an overprediction of the negative shear stress may distort the energy transfer, particularly near the wall, where it may lead to a reduction in the SGS dissipation. From these findings, it can be concluded that although the MSC2 model can represent rotational invariance reasonably well through architectural constraints. However, achieving a higher accuracy requires additional augmentation, such as the inclusion of the STN algorithm, as depicted in the MSC3 model.

\begin{figure*}
    \centering
    \subfloat[\label{eps_0}]{\includegraphics[width=0.4\linewidth]{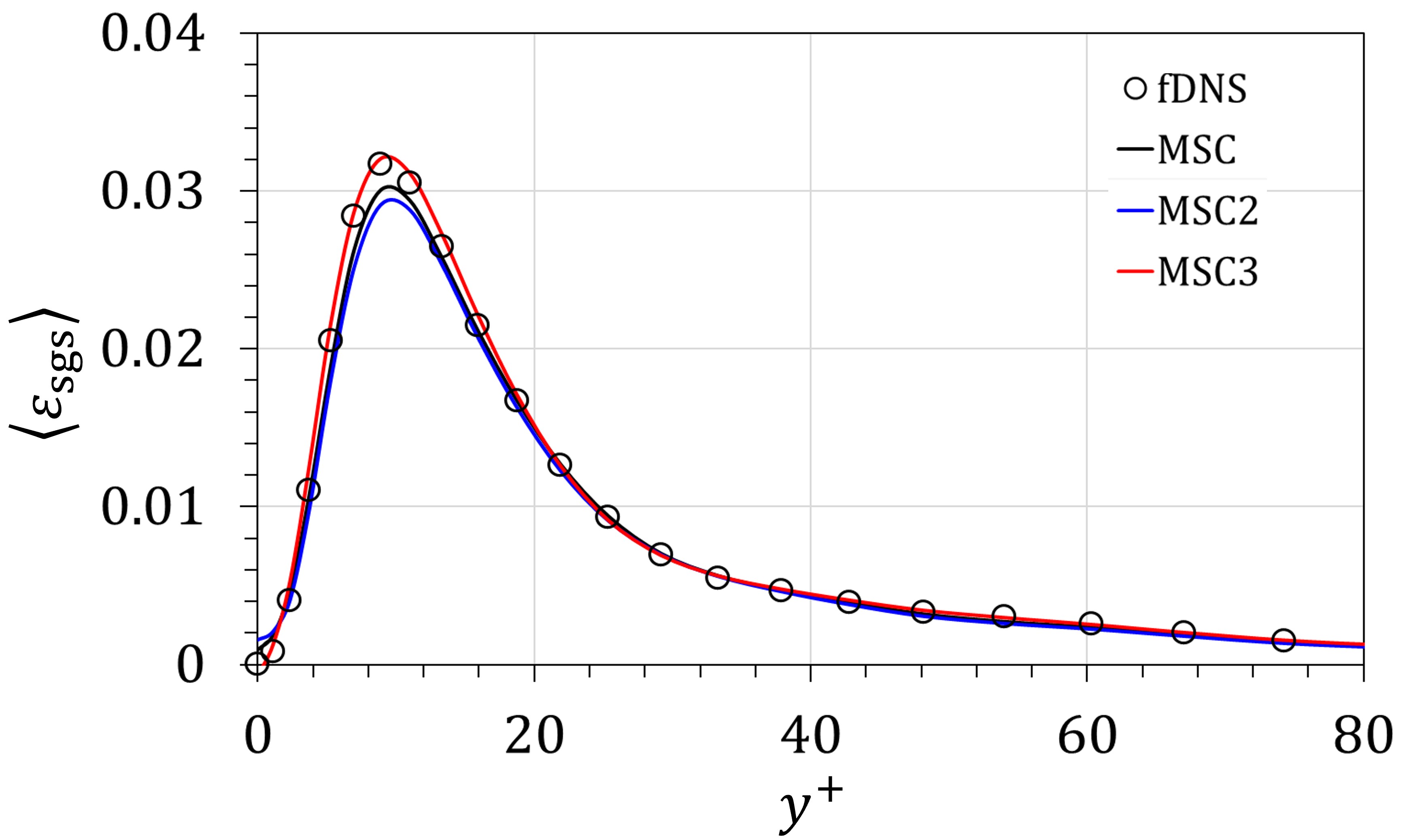}}\hspace{0.01\linewidth}
    \subfloat[\label{eps_90}]{\includegraphics[width=0.4\linewidth]{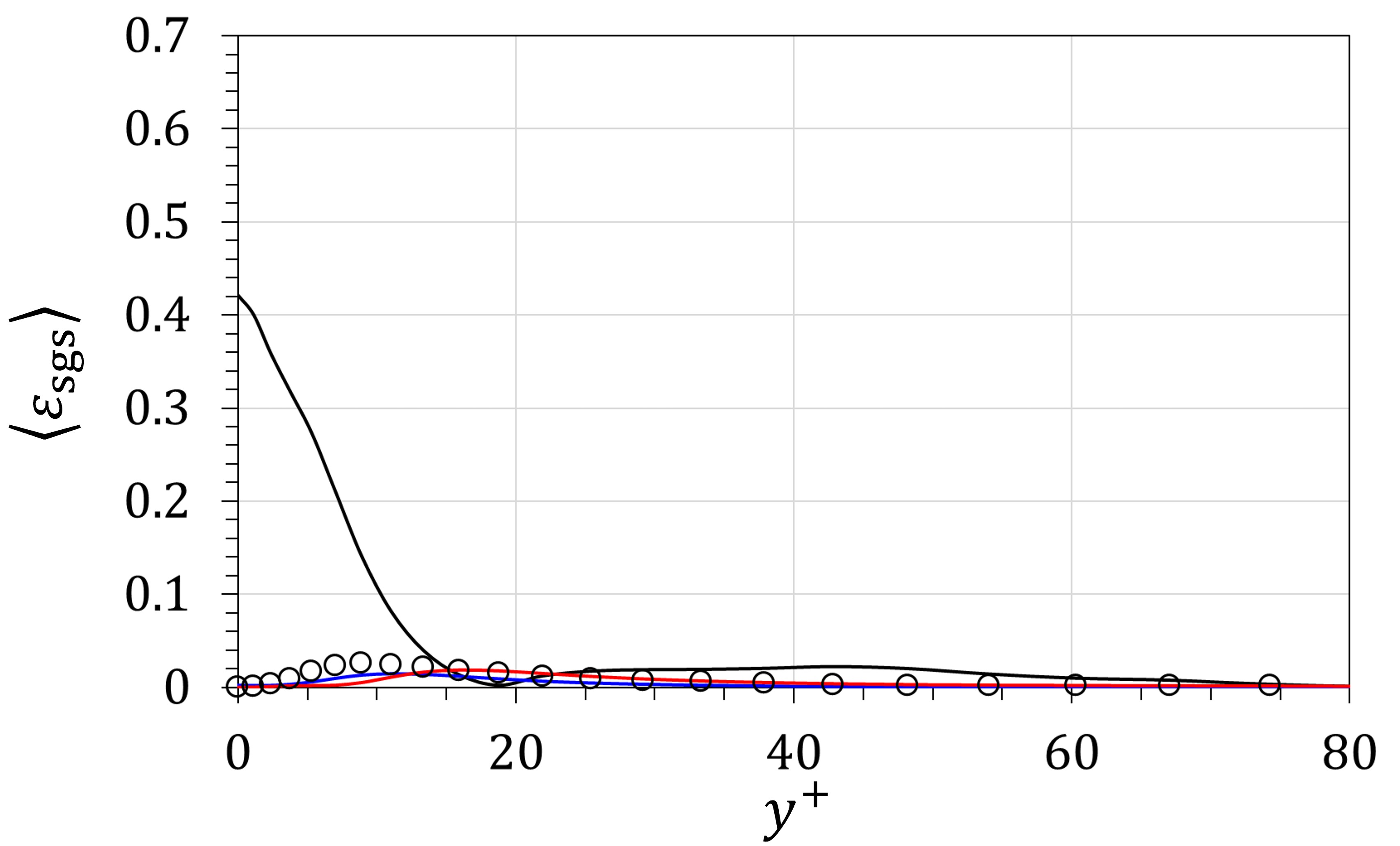}}\hspace{0.01\linewidth}
    \subfloat[\label{eps_270}]{\includegraphics[width=0.4\linewidth]{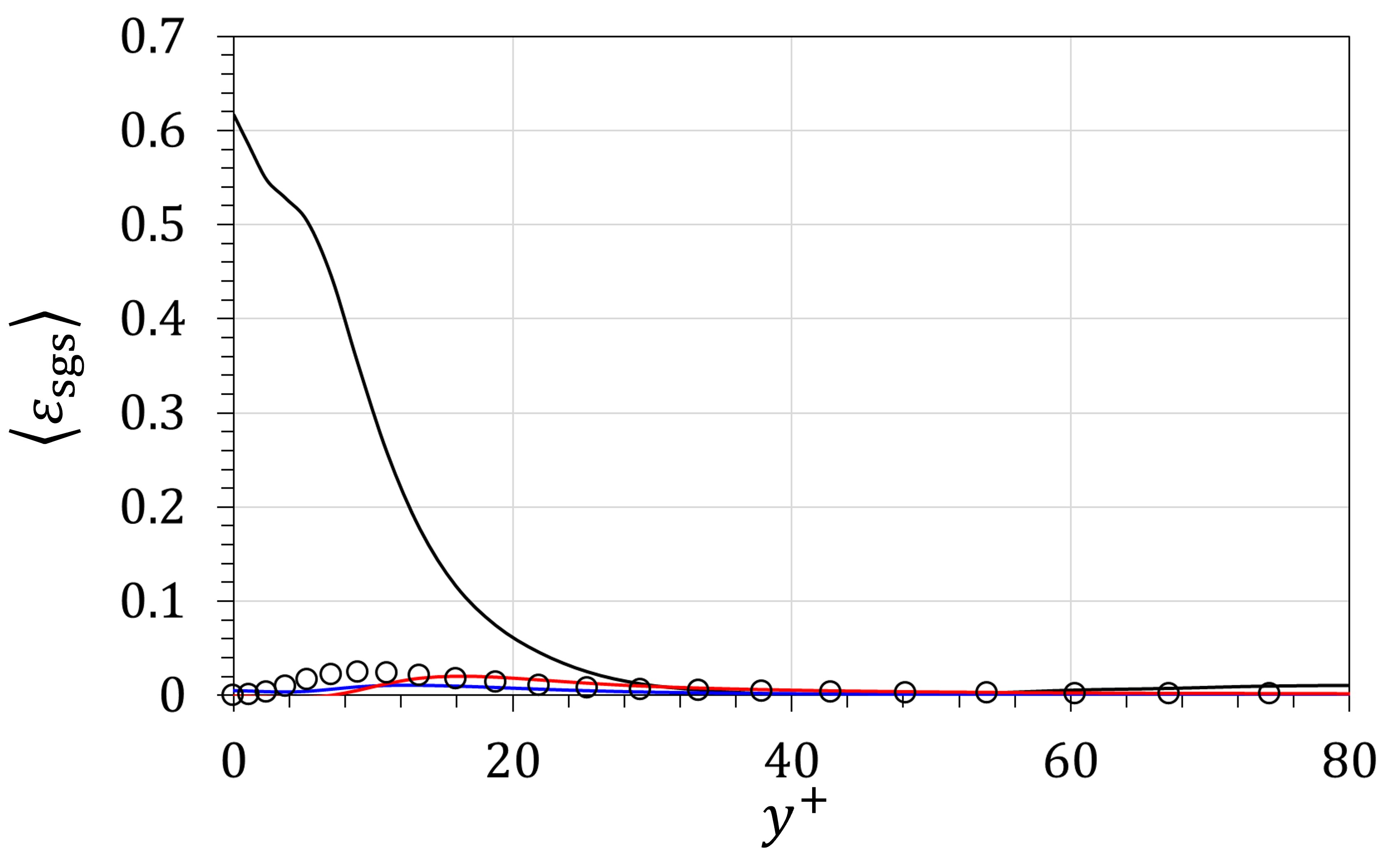}}
    
    \caption{Wall-normal distributions of SGS dissipation $\varepsilon_{\mathrm{sgs}}$ with input rotated by (a) $0^\circ$, (b) $90^\circ$, and (c) $270^\circ$.}
    \label{Figeps}
\end{figure*}

\begin{figure*}
    \centering
    \subfloat[\label{epsb_0}]{\includegraphics[width=0.4\linewidth]{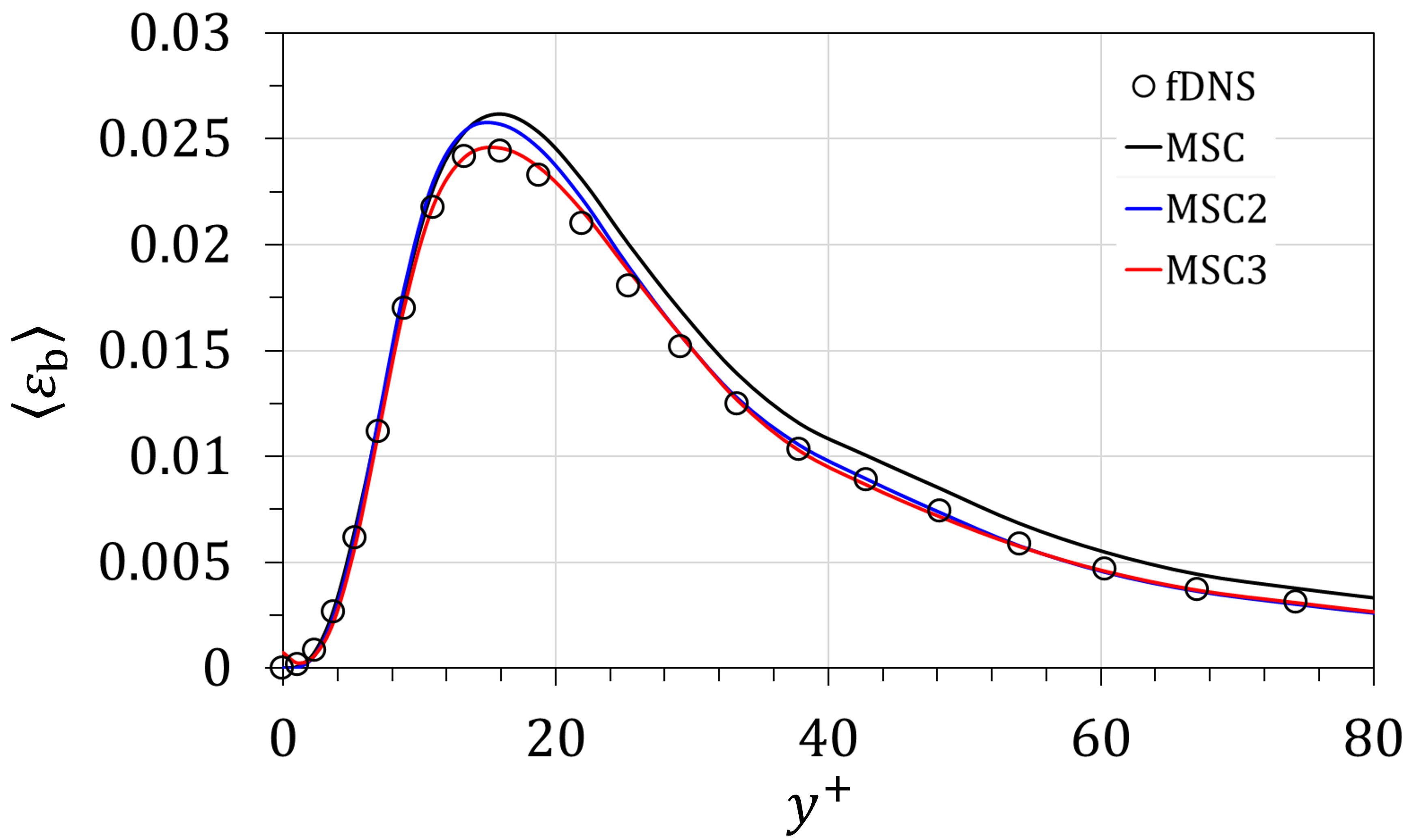}}\hspace{0.01\linewidth}
    \subfloat[\label{epsb_90}]{\includegraphics[width=0.4\linewidth]{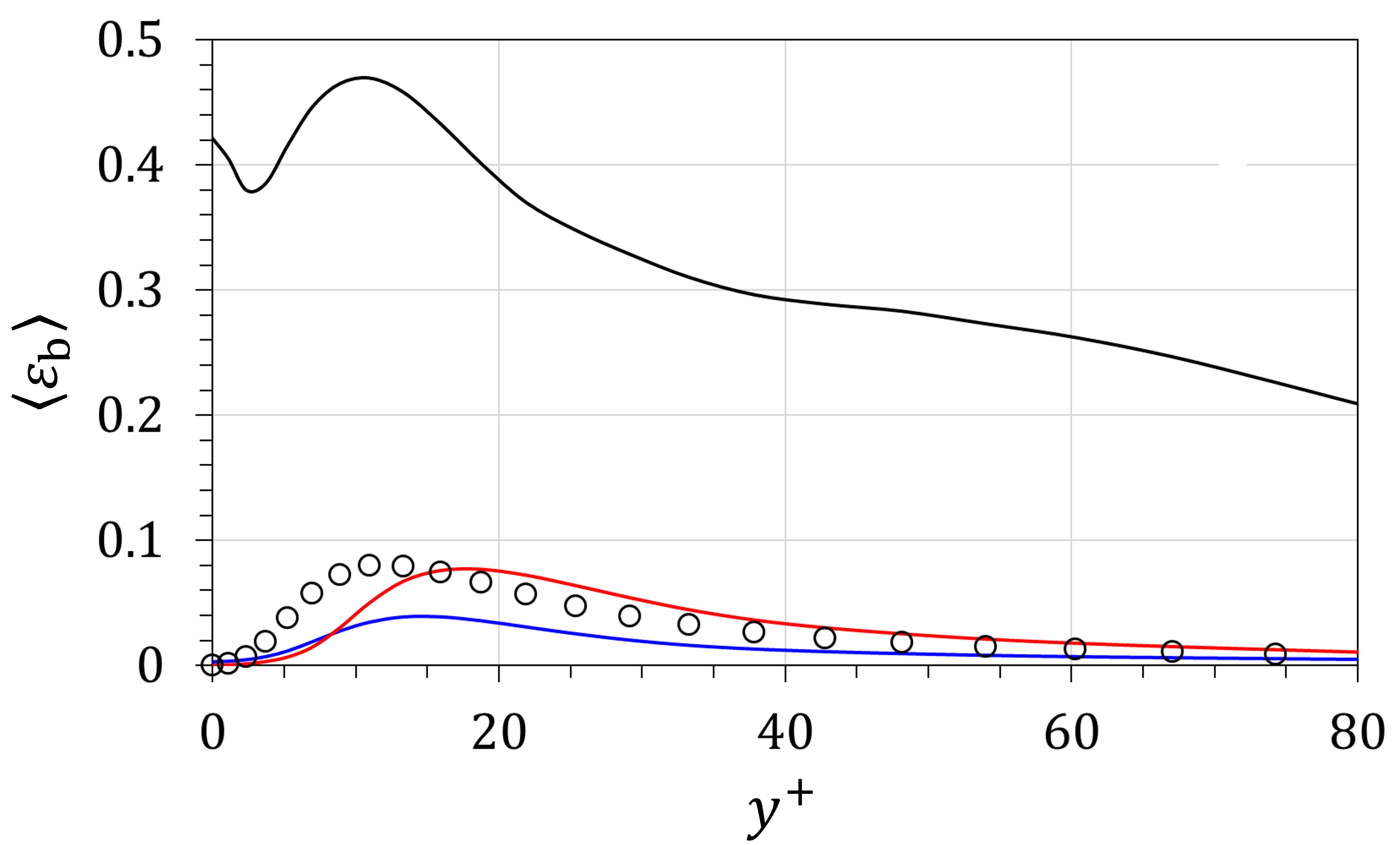}}\hspace{0.01\linewidth}
    \subfloat[\label{epsb_270}]{\includegraphics[width=0.4\linewidth]{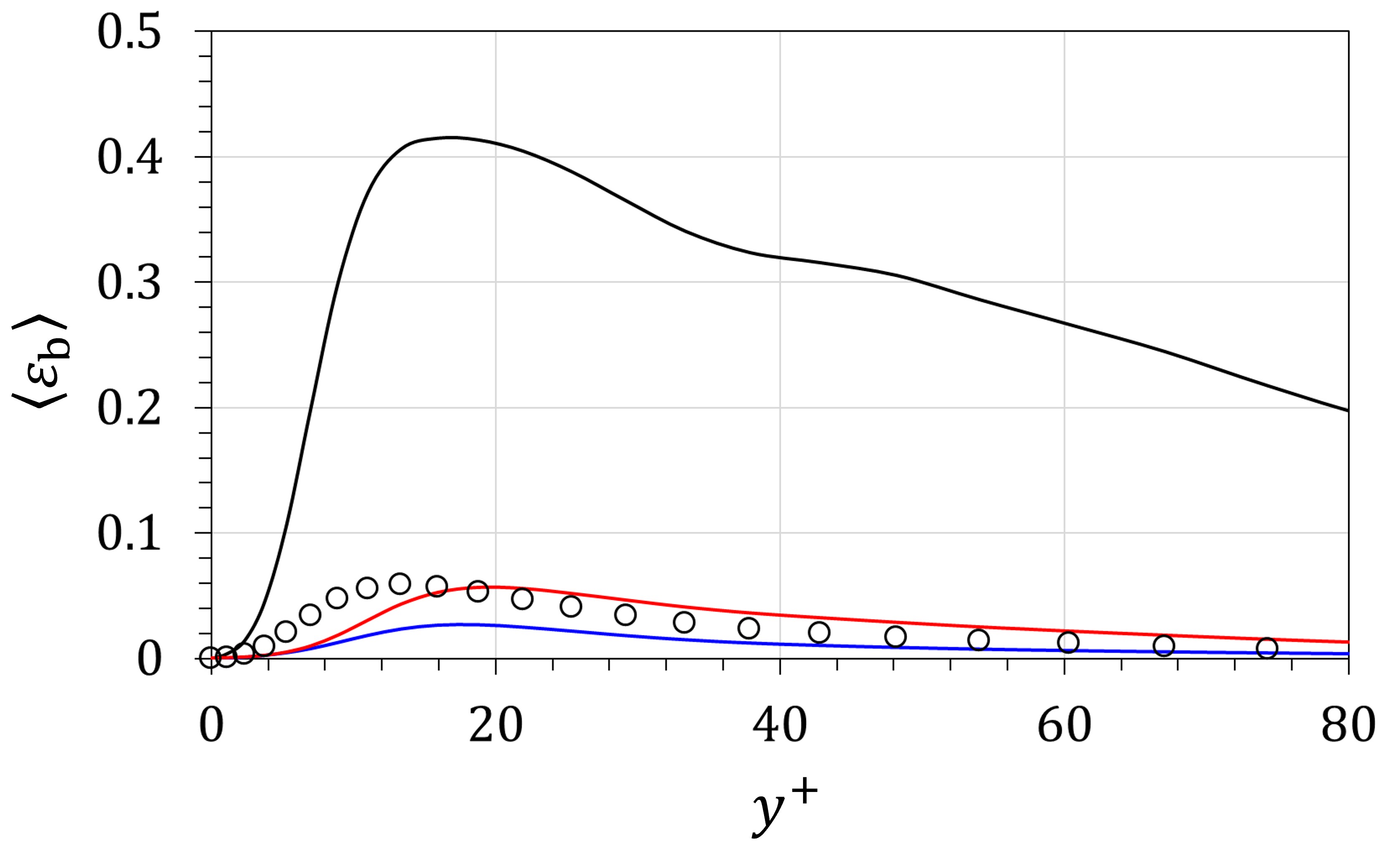}}
    
    \caption{Wall-normal distributions of SGS backscatter $\varepsilon_b$ with input rotated by (a) $0^\circ$, (b) $90^\circ$, and (c) $270^\circ$.}
    \label{Figback}
\end{figure*}

\begin{figure*}
    \centering
    \subfloat[\label{epsb_0}]{\includegraphics[width=0.4\linewidth]{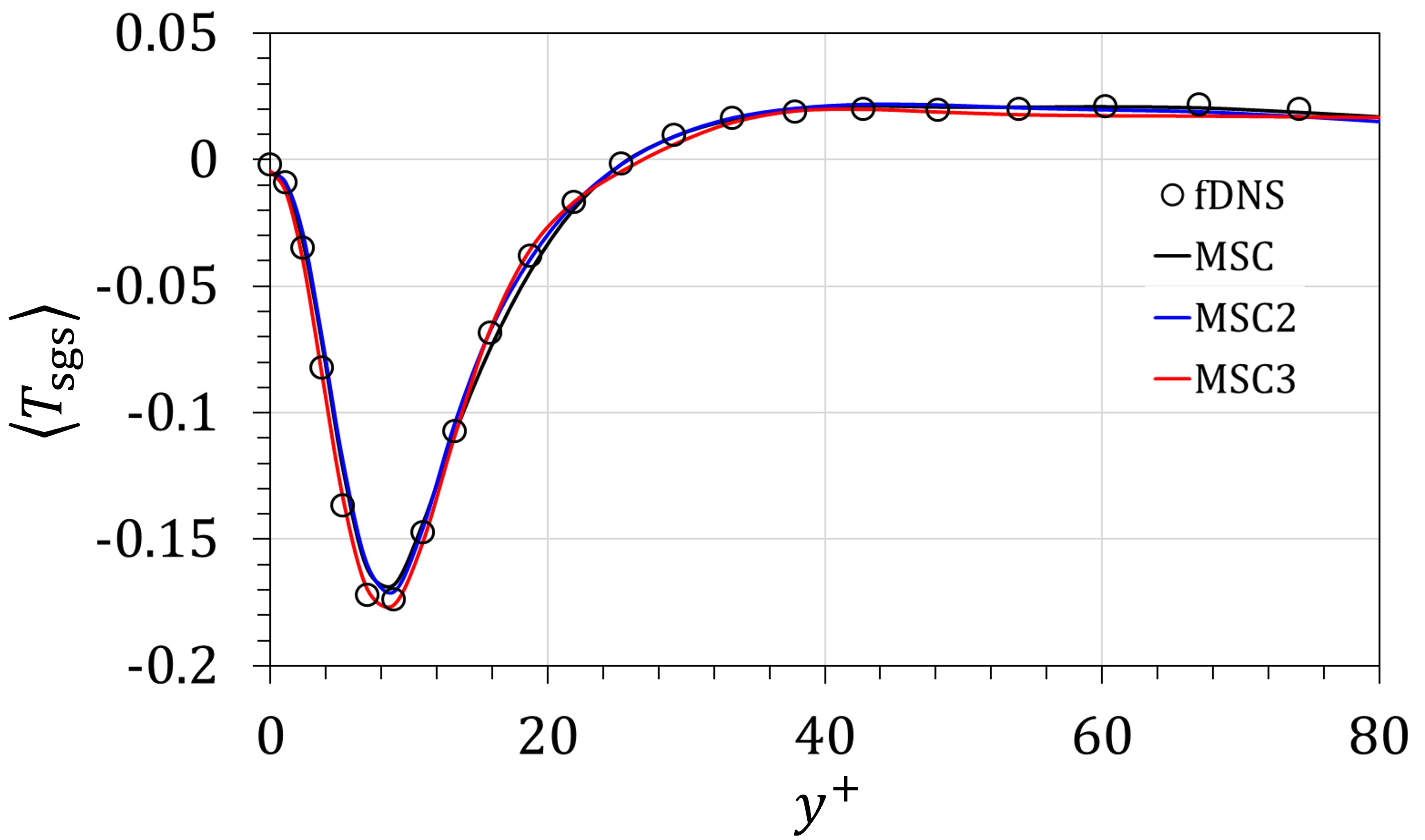}}\hspace{0.01\linewidth}
    \subfloat[\label{epsb_90}]{\includegraphics[width=0.4\linewidth]{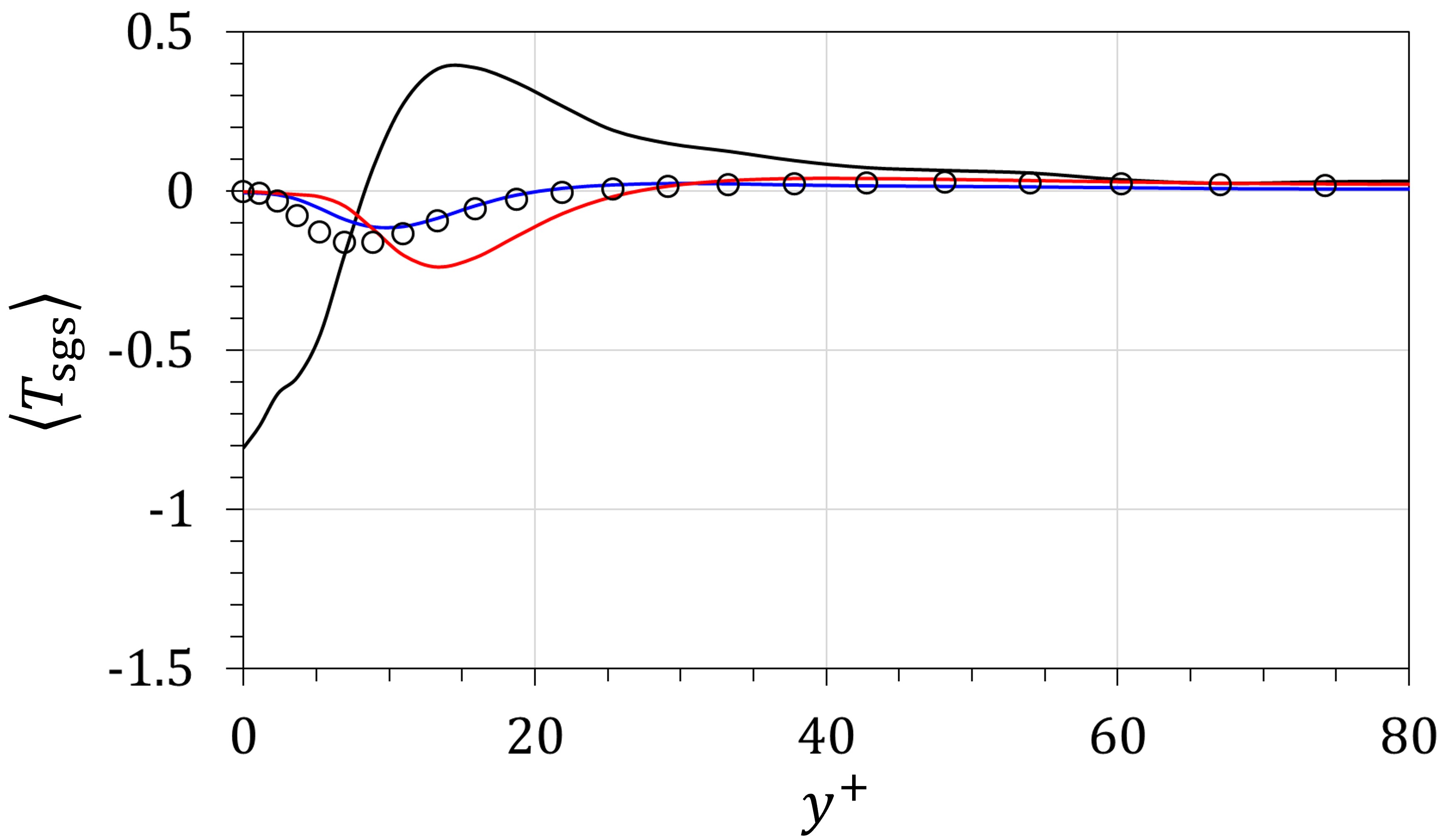}}\hspace{0.01\linewidth}
    \subfloat[\label{epsb_270}]{\includegraphics[width=0.4\linewidth]{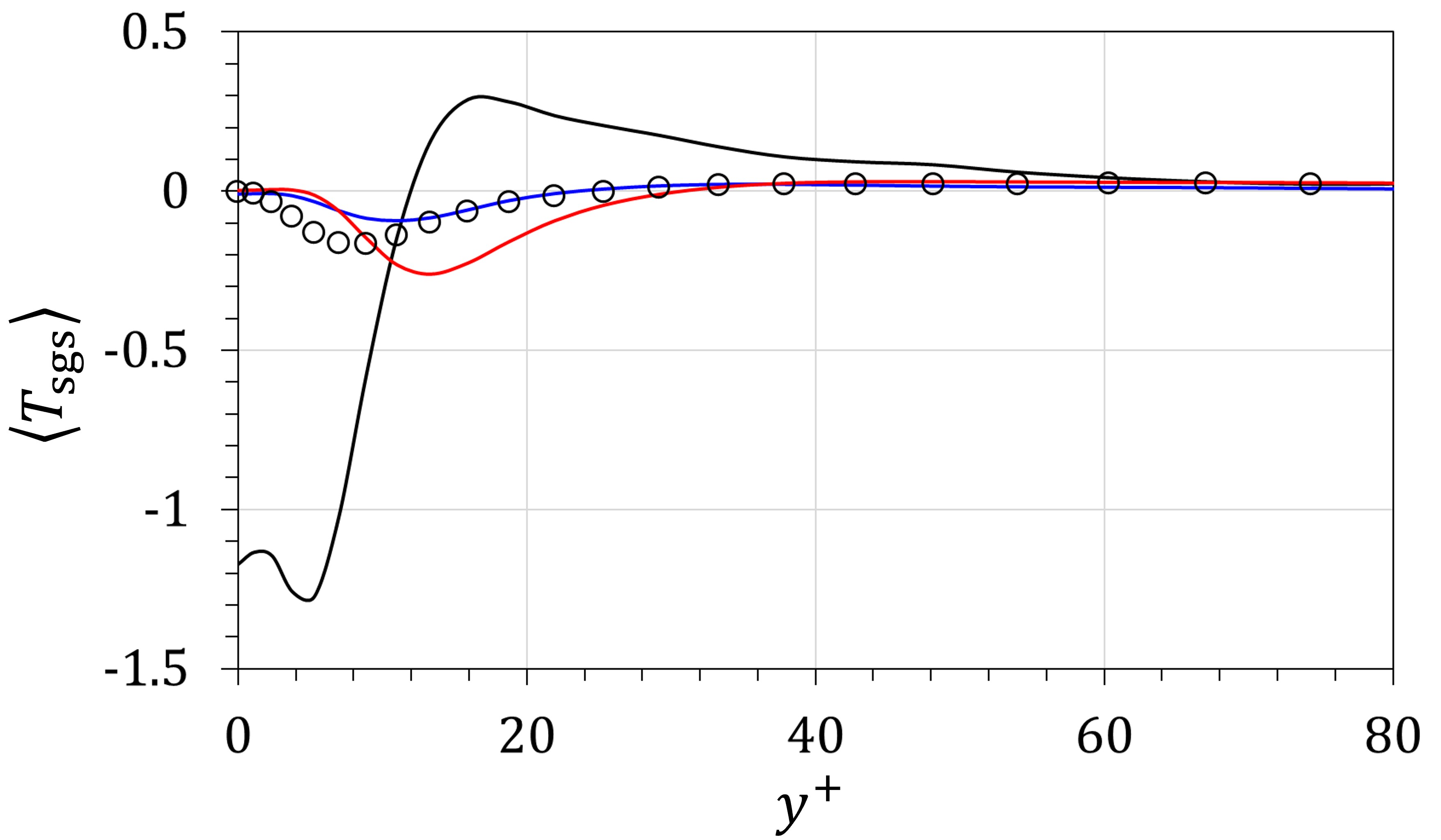}}
    
    \caption{Wall-normal distributions of SGS transport $T_{\mathrm{sgs}}$ with input rotated by (a) $0^\circ$, (b) $90^\circ$, and (c) $270^\circ$.}
    \label{SGSTrans}
\end{figure*}

To further evaluate the data-driven SGS models quantitatively, statistical comparisons are presented to analyzed the underlying physical behavior captured by each model. Following the same procedure described in the preceding section, unseen data with rotated inputs of $90^\circ$ and $270^\circ$ are employed. Fig.~\ref{Figeps} illustrates the SGS dissipation, defined under the local equilibrium assumption as $\varepsilon_{\mathrm{sgs}} = -\tau_{ij} \bar{D}_{ij}$. For a non-rotated input, the MSC and MSC2 models exhibit comparable accuracy, except near the peak region. In contrast, the MSC3 model shows a closer agreement with the fDNS data across the wall-normal direction. Because peak regions are often associated with intense turbulent dynamics and high spatial variability, their accurate prediction is challenging. It is likely that the integration of artificial modifications, particularly the inclusion of the STN algorithm in the MSC3 model, enhances the accuracy of the model in this region. During training, this model was exposed to multiple rotated versions of the input, allowing the STN to infer and correct for orientation differences. This mechanism functions similarly to data augmentation, as discussed by Brener et al.~\cite{brener2024}, but is more efficient~\cite{Jaderberg}, as it does not increase the amount of training data and is integrated directly into the backpropagation process within the data-driven SGS framework. These augmentations enable the model to better capture the turbulent interactions and improve its ability to predict the flow characteristics, even for non-rotated inputs. 

For the rotated inputs of $90^\circ$ and $270^\circ$, the MSC model fails to reproduce the SGS dissipation accurately. This is expected, as this model lacks rotational robustness owing to the presence of bias and BN, as previously discussed. In contrast, the MSC2 and MSC3 models reasonably reproduce the SGS dissipation, although some discrepancies are observed. These errors are mainly attributed to the inaccuracies in the predicted $\tau_{ij}$ and are consistent with the inaccuracies in predicting the number of occurrences, as discussed earlier and shown in Fig.~\ref{bins}.

Figure~\ref{Figback} presents the SGS backscatter, defined as $\varepsilon_b = -\frac{1}{2} \langle \varepsilon_{\mathrm{sgs}} - |\varepsilon_{\mathrm{sgs}}| \rangle$, which represents the reverse transfer of energy from the subgrid scales to resolved scales. Here, the MSC2 and MSC3 models show considerably better alignment with the fDNS data when compared with that of the MSC model in the case of a non-rotated input. The MSC model slightly overpredicts the backscatter, which is likely owing to the spatial distortion introduced by BN. The BN may distort the spatial interaction, which then influences the local interaction on the backward cascade. Despite these slighly overestimated results, all models considered here could produce comparable levels of backscatter, suggesting that modeling the multiscale interactions in the data-driven SGS model based on the MSC model effectively captures the backscatter, as noted by Jalaali $\&$ Okabayashi~\cite{jalaali2025}.

In contrast, for rotated inputs, the MSC model fails to predict the backscatter distribution, exhibiting a severe discrepancy from the fDNS data. Both the MSC2 and MSC3 models demonstrate more robust performance by maintaining a qualitative agreement with the fDNS data, although they slightly underestimate the backscatter, particularly near the wall. Such inaccuracies may lead to inconsistencies in the energy transfer between the resolved and subgrid scales, particularly in the near-wall region where turbulence dynamics are dominant. Despite these inaccuracies, the MSC2 and MSC3 models exhibit the ability to preserve the backscatter distribution even under rotated inputs.

Next, for the SGS transport term in Fig.~\ref{SGSTrans}, defined as $T_{\mathrm{sgs}} = \partial (\tau_{ij} \bar{u}_i)/\partial x_j$, all models perform comparably well for the non-rotated input. The SGS transport term describes the energy transfer within LES systems~\cite{volker2002}, and the accurate prediction of this quantity is crucial. These results show that all models exhibit almost no significant differences in the prediction performance. In the near-wall region ($y^+ < 20$), all models successfully reproduce the negative peak, followed by a nearly zero value away from the wall. These results indicate that the data-driven SGS models preserve the correct near-wall behavior observed in the fDNS data. 

When the input fields are rotated, the MSC model exhibits pronounced deviations, showing an incorrect characteristic of SGS transport. This inconsistency arises from the inability of the model to preserve $\tau_{ij}$ under rotation. In contrast, the MSC2 and MSC3 models maintain the correct overall trend to a considerable extent. Although MSC3 performs better in predicting the SGS dissipation and backscatter, which account for the full SGS contribution, it shows a deviation in SGS transport. This deviation appears to be closely related to the overestimation of $\tau_{ij}$, such as $\tau_{12}$ observed in Fig.~\ref{Fig6}. Because SGS transport involves the divergence of the product of $\tau_{ij}$ and the resolved velocity $\bar{u}_i$, any slight overestimation in $\tau_{ij}$ can affect the transport term, particularly in the regions with strong gradients. Hence, this effect is pronounced for MSC3, with a similar tendency also observed for MSC2.

Overall, the MSC2 and MSC3 models demonstrate good agreement with the fDNS data for both non-rotated and rotated inputs. The modification of the data-driven SGS model does not degrade the predictive accuracy; rather, it enhances the ability of the model to preserve rotational invariance and satisfy material objectivity. The proposed models are not only robust to rotational transformations but also capable of predicting $\tau_{ij}$ with higher accuracy than the baseline MSC model for both rotated and non-rotated inputs. However, it should be noted that these results alone do not guarantee accuracy or stability in actual LES computations. This study neglected the effects of numerical errors and cumulative propagation of uncertainties during the temporal evolution of the flow. It also assumed that $\bar{D}_{ij}$ is completely invariant under rotation. Within the full LES framework, the computation incorporates the dynamic evolution of the flow in a rotating frame and other numerical complexities. Verifying whether these advantages persist in the full LES environment remains an essential next step, as the interaction between the modeled and resolved scales introduces additional complexities that are not captured in the current analysis.

\section{Conclusion}
To ensure physical consistency, a data-driven SGS model must satisfy the invariance principles underlying turbulence modeling or the principle of material objectivity, as pointed out by Spalart~\cite{spalartphilosophies2015}. Therefore, in this study, we proposed a novel data-driven SGS model for LES, with a focus on enforcing rotational invariance for a wall-bounded flow. Building upon the multiscale CNN-based SGS framework proposed by Jalaali \& Okabayashi~\cite{jalaali2025} (MSC model), we introduced two novel data-driven SGS models, denoted as the MSC2 and MSC3 models in this paper. The MSC2 model enforced material objectivity by excluding the bias terms and BN layers, whereas the MSC3 model further incorporated an STN algorithm\cite{Jaderberg} to learn the spatial alignment under the rotated input.

We conducted \textit{a priori} assessments to demonstrate the performance of the proposed data-driven SGS models. We evaluated the predictive capability and physical consistency of the data-driven SGS models using unseen data that differed from the training set. Overall, the results showed that both MSC2 and MSC3 outperformed the baseline MSC model in predicting the SGS quantities, including $\tau_{ij}$, dissipation, backscatter, and SGS transport. This indicates that the models successfully captured the energy transfer processes between the resolved and subgrid scales. Although all models performed well on non-rotated inputs, the baseline MSC model exhibited a strong orientation dependence under rotated conditions. This was likely because of the inclusion of bias and BN, which caused the data-driven SGS model to be fitted with the training data distribution. In particular, BN normalized the input features across mini-batches, which could have disrupted the spatial interactions within the flow field. Because turbulent flows are influenced by vortex interactions, preserving these spatial correlations is essential for constructing physically consistent data-driven SGS models. In addition, the exclusion of bias and BN in the MSC2 and MSC3 models allowed them to maintain spatial representations and vortex interactions of the input feature $\bar{D}_{ij}$. Nevertheless, the integration of the STN algorithm in the MSC3 model helped the model recognize and correct for input orientation, thereby improving both its accuracy and robustness. Although the MSC3 model achieved the highest overall accuracy for both non-rotated and rotated input cases among all models, a noticeable limitation was observed in the number of occurrences of $\tau_{12}$, where the negative value was overpredicted. This tendency likely originated from insufficient rotated samples during STN training, leading to partial misrepresentation of the shear structures.  

However, the present evaluation was carried out only in \textit{a priori} test under simplified conditions by excluding numerical errors and temporal evolution of the flow in the LES framework. Future studies are expected to assess the model performance in the LES framework, where model-resolved interaction would play a more dominant role. Extending the investigation to more complex flow configurations and broader training datasets is expected to further improve the generalization and robustness of the proposed models. This model could be applied to rotational flows, such as those found in turbomachinery, for which, to the best of the authors’ knowledge, no existing data-driven SGS model has yet been developed.

Although the present implementation was limited to canonical wall-bounded turbulence and additional investigations are required to fully explore this potential, this study represented a step forward in developing data-driven SGS models that satisfied rotational invariance. The proposed data-driven SGS models are identified as a promising direction for developing SGS models consistent with physical constraints. They demonstrate the potential of the multiscale CNN-based SGS framework to satisfy the principle of material objectivity in machine-learning turbulence models, and are intended to bridge the gap between purely data-driven approaches and traditional physics-based SGS formulations.

\begin{acknowledgments}
This work was partly supported by the Research Proposal-based Use of the Project for Nurturing Student Competing with the World at the Large-Scale Computer System $\--$ D3 Center, the University of Osaka. This study was financially supported by JSPS KAKENHI grant No. JP25K07584 and Indonesian Education Scholarship Program (BPI), administered by the Center for Higher Education Funding and Assessment (PPAPT), Ministry of Higher Education, Science, and Technology of the Republic of Indonesia and the Indonesia Endowment Fund for Education (LPDP), Ministry of Finance of the Republic of Indonesia.
\end{acknowledgments}

\appendix

    \begin{table*}[ht]
    \centering
    \small 
    \renewcommand{\arraystretch}{1.5}
    \caption{Comparison of data-driven SGS models.}
    \begin{tabular}{lccccccc}
        \toprule
        \text{Parameter} & MSC\cite{jalaali2025} & MSCa & MSCb & MSCc & MSC2 & MSC3  \\
        \hline
        Exclusion of bias & no & yes & no  & no  & yes & yes \\
        Exclusion of BN   & no & no  & yes & no  & yes & yes \\
        STN               & no & no  & no  & yes & no  & yes \\
        \toprule
    \end{tabular}
    \label{table:appndA}
    \end{table*}

\begin{figure*}
    \centering
    \subfloat[\label{append_0}]{\includegraphics[width=0.4\linewidth]{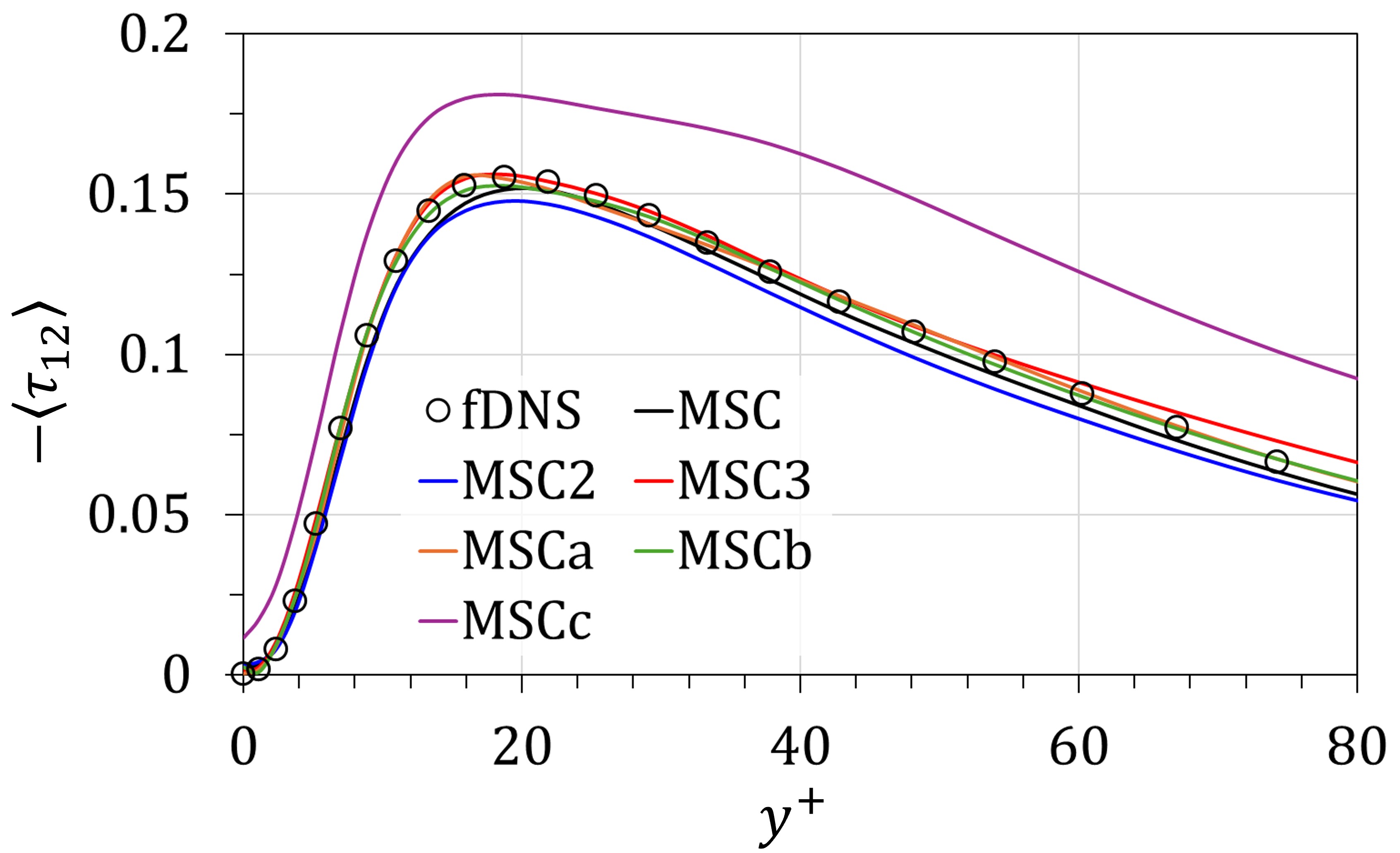}}\hspace{0.01\linewidth}
    \subfloat[\label{append_90}]{\includegraphics[width=0.4\linewidth]{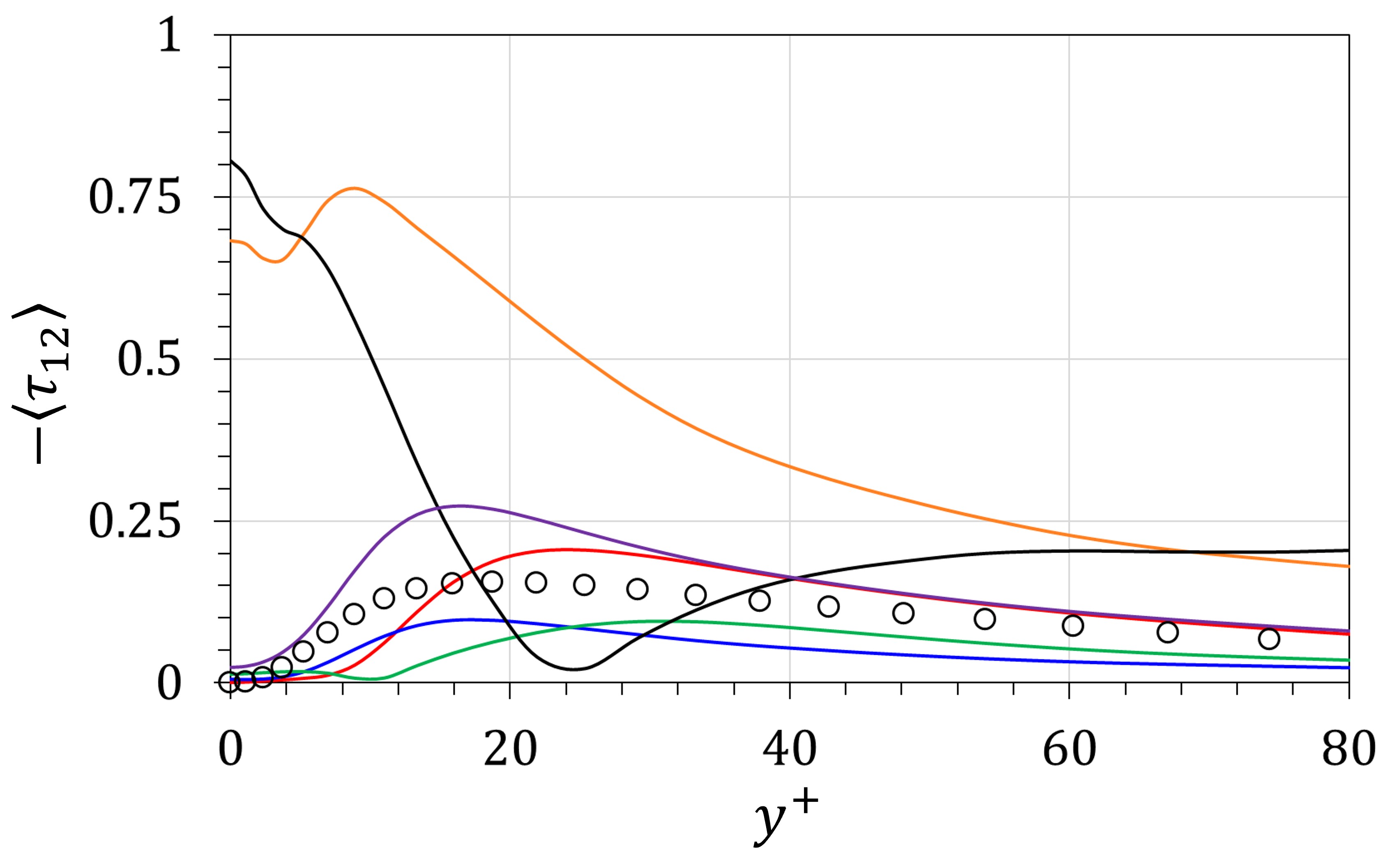}}\hspace{0.01\linewidth}
    \subfloat[\label{append_270}]{\includegraphics[width=0.4\linewidth]{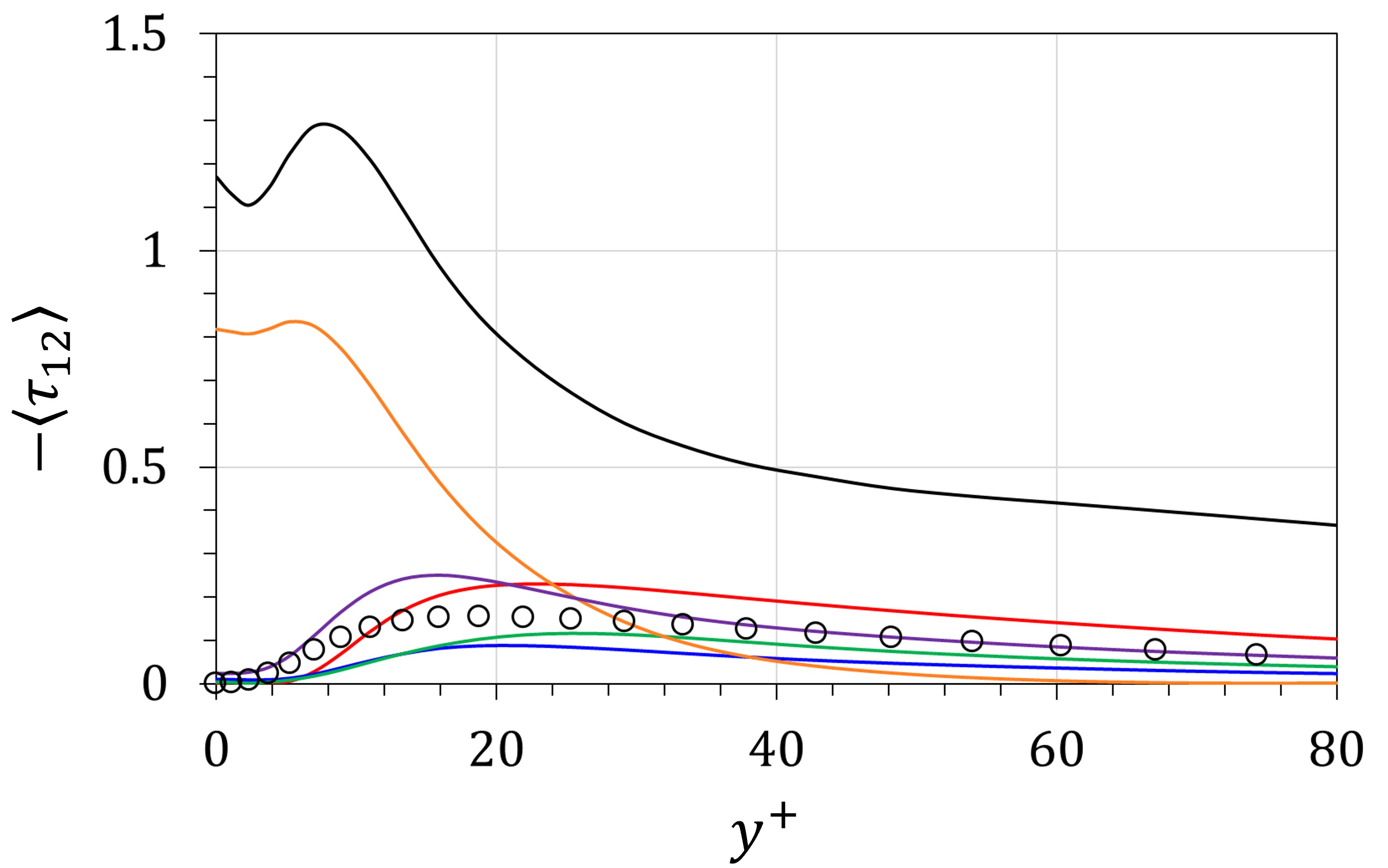}}
    
    \caption{Wall-normal distributions of $\tau_{12}$ with input rotated by (a) $0^\circ$, (b) $90^\circ$ and (c) $270^\circ$.}
    \label{FigApx1}
\end{figure*}

\section{Assessment of the effect of bias, BN, and STN algorithm on the objectivity in \textit{a priori} test.}
In this section, we report the effects of excluding the bias and batch normalization (BN), as well as introducing the spatial transformer network (STN) algorithm\cite{Jaderberg} in a data-driven SGS model, as determined through \textit{a priori} tests. The exclusion of bias and BN has been shown to improve the objectivity of data-driven SGS models, yet their respective contributions to this improvement remain unclear. Therefore, a dedicated evaluation of these architectural components is essential to identify the architectural components of the DNN that most significantly influence the ability of the data-driven SGS model to accurately predict $\tau_{ij}$ and infer the objectivity. Here, all models were trained using the same dataset (i.e., the total number of training datasets was kept identical across all models, with MSCc and MSC3 employing rotated data, as described in Section~II) and training procedure, ensuring a fair and consistent comparison. The model configurations used in this study are summarized in Table~\ref{table:appndA}. To evaluate both the predictive performance and rotational invariance of the models, we examined the wall-normal distribution of the shear stress component $\tau_{12}$. 

\textit{A priori} test results for the wall-normal distribution of $\tau_{12}$ are shown in Fig.~\ref{FigApx1}. It can be seen that all models exhibit good agreement for the non-rotated input, except for the MSCc model. This inaccuracy is likely owing to a conflict between the architectural components. Although the STN is designed to promote spatial alignment and rotational invariance by learning and correcting the input transformations, the inclusion of bias and BN may undermine this capability. BN performs normalization across mini-batches, which can disrupt the spatial coherence, whereas the bias term introduces constant offsets that may shift the learned data distribution. As a result, the extracted features in the MSCc model become inconsistent, leading to reduced accuracy. Although the MSCc model shows degraded performance on non-rotated inputs, it exhibits improved accuracy when tested with rotated inputs. This indicates that the STN module retains its robustness in detecting and correcting the spatial transformations, despite interference from bias and BN. In comparison with the baseline model, that is the MSC model, the introduction of the STN algorithm enables the MSCc model to learn rotational features. 

For the MSCa model, the exclusion of bias does not produce any significant effect on the non-rotated input, as it preserves its predictive accuracy. However, it exhibits noticeable deviations when tested under the rotated input. This result is expected, as the presence of BN alters the structure of the spatial correlations. By normalizing the output of the CNN layers across mini-batches, BN implicitly partitions the training data. We argue that this partitioning breaks the spatial coherence and disturbs the underlying vortex interaction, which can lead to unphysical representations. Turbulent flows are closely related to the local interaction and spatial coherence\cite{davidson2015}, and the exclusion of BN is critical when constructing data-driven SGS models. In addition, in the MSC model, alteration in the flow representation owing to BN appears to be corrected by bias. This result is consistent with the observations from the MSCb model, where the exclusion of BN enables it to better predict the rotated input. Because the input variable $\bar{D}_{ij}$ is inherently objective, the MSCb model benefits from this property and is able to develop a more rotation-invariant representation. However, the inclusion of bias in the MSCb model reintroduces an offset into the network, which leads to a degradation in accuracy and failure to satisfy the material objectivity requirement when tested under rotated input. The behaviors of the MSC2 and MSC3 models are discussed in the main section, where their performances under both rotated and non-rotated conditions have been analyzed in detail. Therefore, further discussion of the  MSC2 and MSC3 models is omitted here.

Thus, it becomes clear that the exclusion of bias and BN promotes rotational invariance by preserving the spatial interaction between the input features. Bias introduces fixed offsets, whereas BN disrupts the spatial correlations through mini-batch normalization, which can compromise the ability of the model to maintain objectivity. However, achieving invariance alone may not be sufficient for an accurate prediction despite the use of the objective input variable $\bar{D}_{ij}$. The integration of the STN algorithm equips the data-driven SGS model with the ability to recognize input orientation issues and correct them. STN thereby enhances both the objectivity and robustness of the data-driven SGS model, especially under rotated conditions. Hence, while removing bias and BN helps enforce invariance, incorporating STN is crucial for enabling the model to learn rotational representations and achieve physically consistent predictions.

\end{CJK}
\renewcommand{\bibsection}{\section*{REFERENCES}}
\bibliographystyle{aipnum4-2}
\bibliography{Reference}

\end{document}